\documentclass[a4paper,11pt]{article}  
\usepackage{jheppub}
\usepackage{epsfig}
\usepackage{color}
\usepackage{latexsym}
\usepackage{amsmath}
\usepackage{amssymb}
\usepackage{dsfont}
\usepackage{verbatim}
\usepackage{bm}
\usepackage{multirow}

\newcommand{\I}{\mathrm{i}}

\begin{document}

\title{Vector and scalar charmonium resonances with lattice QCD}

\author[a]{C. B. Lang,}
\author[b]{Luka Leskovec,}
\author[c]{Daniel Mohler}
\author[b,d,e]{and Sasa Prelovsek}
\affiliation[a]{Institute of Physics, University of Graz,\\
Universit\"atsplatz 3, A--8010 Graz, Austria}
\affiliation[b]{Jozef Stefan Institute,\\
Jamova 39, 1000 Ljubljana, Slovenia}
\affiliation[c]{Fermi National Accelerator Laboratory,\\
P.O. Box 500, Batavia, Illinois 60510-5011, U.S.A.}
\affiliation[d]{Department of Physics, University of Ljubljana,\\
Jadranska 19, 1000 Ljubljana, Slovenia}
\affiliation[e]{Theory Center, Jefferson Lab,\\
12000 Jefferson Avenue, Newport News, VA 23606, USA}

\emailAdd{christian.lang@uni-graz.at}
\emailAdd{luka.leskovec@ijs.si}
\emailAdd{dmohler@fnal.gov}
\emailAdd{sasa.prelovsek@ijs.si}

\abstract{
We perform an exploratory lattice QCD simulation of $D\bar D$ scattering, aimed at determining the masses as well as the decay widths of charmonium resonances above open charm threshold. Neglecting coupling to other channels, the resulting phase shift for $D\bar D$ scattering in $p$-wave yields the well-known vector resonance $\psi(3770)$. For  $m_\pi\!=\!156~$MeV, the extracted resonance mass and the decay width  agree with experiment within large statistical uncertainty.  The scalar charmonium resonances present a puzzle, since only the ground state $\chi_{c0}(1P)$ is well understood, while there is no commonly accepted candidate for its first excitation. We simulate $D\bar D$ scattering in $s$-wave   in order to shed light on this puzzle. The resulting phase shift supports the existence of a yet-unobserved narrow resonance with a mass slightly below $4~$GeV. A scenario with this narrow resonance and a pole at $\chi_{c0}(1P)$ agrees  with the energy-dependence of our phase shift. Further lattice QCD simulations and experimental efforts are needed to resolve the puzzle of the excited scalar charmonia.}


\arxivnumber{}

\maketitle

\section{Introduction}\label{sec_introduction}
Charmonia $\bar cc$ well below open-charm threshold $\bar DD$ are among the best understood hadrons.  Their spectra and selected transition matrix elements are successfully described by lattice QCD simulations and QCD motivated models. Recent lattice calculations  have performed the necessary extrapolations and  considered spectra  \cite{Mohler:2014ksa,Galloway:2014tta} as well as certain radiative transitions \cite{Becirevic:2014rda,Donald:2012ga}. For states well-below open charm  threshold, the main remaining uncertainty  is the neglect of charm-annihilation Wick contractions in lattice simulations. 

The most interesting charmonium and charmonium-like states lie near or above open charm thresholds. During the past decade a plethora of states that can likely not be interpreted as conventional $\bar cc$ have been discovered in experiment (for a review see for example \cite{Brambilla:2014jmp,Esposito:2014rxa}).  These states have been treated theoretically making simplifying assumptions and reliable quantitative results for those hadrons are not available. In particular, all of the lattice simulations so far have ignored the strong decay of the charmonium resonances to  a pair charmed mesons $\bar cc \to \bar D^{(*)}D^{(*)}$, which typically represents the main decay mode. Except in a few simulations \cite{Bali:2011rd,Ozaki:2012ce,Prelovsek:2013cra,Lee:2014uta}, the effect of the threshold on the  near-threshold states has been neglected.  The most extensive spectrum of charmonia has been obtained in simulations with $N_f=2+1$ dynamical flavors at $m_\pi\simeq 400$  MeV  \cite{Liu:2012ze}, but the determination neglects the unstable nature of the states and relies on extracting the energy levels with only quark-antiquark interpolating fields, which may lead to unphysical results close to multi-hadron thresholds \cite{Lang:2011mn,Dudek:2012xn,Mohler:2012na}. 
 
Here we present a  lattice QCD simulation of the vector ($J^{PC}=1^{--}$) and scalar  ($0^{++}$)  charmonium resonances above $\bar DD$ threshold, taking into account their strong decay to $\bar DD$. The lowest vector resonance above open charm threshold, the $\psi(3770)$ is well established in experiment \cite{pdg14}, and we extract its width by simulating $\bar DD$ scattering in $p$-wave. In contrast to that, the experimental and theoretical status of scalar charmonia is puzzling: the only well-established state is the  ground state $\chi_{c0}(1P)$, while there is no commonly accepted candidate for its first excitation $\chi_{c0}(2P)$. We present a study of $\bar D D$   scattering in $s$-wave, aiming to address this open problem. We also consider possible effects of the $\bar DD$ threshold on the vector $\psi(2S)$ and scalar $\chi_{c0}$ charmonia, which lie below threshold. 

To study the hadrons present in these two channels our analysis makes a number of simplifying assumptions based on phenomenology, model calculations and experimental data:

\begin{itemize}
\item We only include interpolating fields of a quark-antiquark and meson-meson type. For the meson meson interpolators we restrict our study to $\bar{D}D$ interpolators for the $\Psi(3770)$,  and  $\bar{D}D ~\&~ J/\Psi\omega$ interpolators for $\chi_{c0}^\prime$.
\item We assume that elastic decay into $\bar{D}D$ is a good approximation for extracting the mass and width of the states. For the $\Psi(3770)$, where $Br^{exp}[\psi(3770)\to D\bar D]=93\pm 9\%$, we neglect \emph{all} further possible two- and more hadron channels, in particular decay into light hadrons through charm annihilation diagrams, $J/\Psi\pi\pi$, $J/\Psi\eta$ and $\eta_c\omega$. Possible influence of  thresholds $\chi_{c0}\omega$, $\chi_{c1}\omega$ and $\bar{D}D^*$, which lie above $\psi(3770)$, is also omitted. For the study of the scalar channel we assume elastic scattering of $\bar D D$ and neglect all other open channels such as the $\eta_c \eta$ or $\chi_{c1} \eta$. Note that $\eta$ is a flavour singlet on our $N_f=2$ ensemble, and is therefore relatively heavy.
\item In the scattering analysis of the discrete energy levels we model the scattering amplitude using various model forms. For the $\Psi(3770)$ our model assumptions are based on phenomenology (experiment branching fractions and upper limits), while they are based on potential model expectations in the case of the $\chi_{c0}^\prime$.
\end{itemize}

\section{Open questions for charmonia of interest} \label{sec:open_questions}
 
\subsection{Vector charmonia}
 
The $\psi(3770)$ with $M=3773.15\pm 0.33~$MeV and $\Gamma=27.2\pm 1.0~$MeV is located only $\simeq 45~$MeV above  $\bar DD$ threshold \cite{pdg14,Anashin:2011kq}. We focus here on its dominant decay mode $\psi(3770)\to \bar DD$ in $p$-wave with branching fraction $0.93 \genfrac{}{}{0pt}{}{+8}{-9}$ \cite{pdg14}.   It is a well-established experimental resonance and is generally accepted to be predominantly the conventional $^{2s+1}nL_J=^{3\!}1D_{1}$ $\bar cc$ state \cite{Rosner:2001nm,Rosner:2004wy,Eichten:2007qx}. There is an ongoing discrepancy between results from BES-II \cite{Ablikim:2008zzb} and Cleo \cite{Besson:2005hm} regarding the non-$\bar{D}D$ part of the branching fraction which may be connected to neglecting interference effects in the BES-II analysis. Significant non-$\bar{D}D$ decays into light hadrons can occur if there is non-negligible mixing with the $\psi(2S)$ \cite{Wang:2004kf}. For our analysis we neglect disconnected contributions that would cause decay into light hadrons and treat the decay into $\bar{D}D$ as elastic, neglecting the decays into $J/\psi\,\pi\pi$ and $J/\psi\,\eta$ that have tiny branching fractions \cite{pdg14}. Our aim is to perform a determination of the $\psi(3770)$ resonance mass and $\psi(3770)\to \bar DD$ decay width using a lattice simulation for $\bar DD$ scattering in $p$-wave. 
 
We also investigate  whether the $\bar D D$ threshold has any effect on $\psi(2S)$, which is the first radial excitation of $J/\psi$ and is  situated  $\simeq 42~$MeV below threshold. Such a possibility was discussed in relation to the Fermilab-MILC preliminary results \cite{DeTar:2012xk} where a simple analysis of the spin-averaged $2S$ state appeared high with respect to experiment, although large systematic uncertainties related to excited state contaminations were observed. A more recent HPQCD study \cite{Galloway:2014tta} finds no significant discrepancy. The mixing of the vector charmonia with a pair of two charmed mesons was first simulated in \cite{Bali:2011rd}, where only $D_1\bar D$ in $s$-wave was considered and the width of $\psi(3770)$ was not extracted.    
 
\subsection{Scalar charmonia}
 
The only well established scalar charmonium state is the ground state $\chi_{c0}(1P)$, interpreted as the  $^31P_0$   $\bar cc$  and located well below the open charm threshold. A further known resonance, the $X(3915)$ with $\Gamma=20\pm 5~$MeV is seen only in $J/\psi\, \omega$ and $\gamma\gamma$  decay channels \cite{pdg14}. BaBar has determined its $J^P$ quantum numbers to be $0^+$ \cite{Lees:2012xs} which would only allow $J^{PC}=0^{++}$.  This spin-parity determination by BaBar assumes that a $J^P=2^+$ resonance would be produced in the helicity 2 state, which might not be justified for an exotic meson\footnote{For arguments in favor of the $X(3915)$ as a $J=2$ resonance see Ref. \cite{Zhou:2015uva}.} \cite{Brambilla:2014jmp}. As a consequence, the PDG recently assigned  $X(3915)$  to be  $\chi_{c0}(2P)$ \cite{pdg14}, but a number of convincing reasons given by Guo \& Meissner \cite{Guo:2012tv} and Olsen \cite{Olsen:2014maa} raise serious doubts about this assignment:
\begin{itemize}
\item
 The dominant decay mode of scalar charmonium above open charm threshold is expected to be a "fall-apart" mode into $\bar D D$   that would lead to a relatively broad resonance. In particular the width into $\bar D D$ is expected to be much larger than for the well-established $\chi_{c2}(2P)$ \cite{pdg14}, which decays to $\bar D D$ in d-wave. Yet $m_{D\bar D}$ invariant mass spectra of  several experiments show no evidence for  $X(3915)\to D\bar D$.  This also indicates that the  $\bar DD$ width extracted from the present lattice simulation cannot be compared to $X(3915)$.
\item
 The  spin-splitting $m_{\chi_{c2}(2P)}-m_{\chi_{c0}(2P)}$ within this assignment seems too small  compared to $m_{\chi_{b2}(2P)}-m_{\chi_{b0}(2P)}$ or $m_{\chi_{c2}(1P)}-m_{\chi_{c0}(1P)}$.  
\item The partial width for the OZI suppressed $X(3915)\to \omega J/\psi$ seems too large \cite{Guo:2012tv}, which is translated to two contradicting limits for this decay in \cite{Olsen:2014maa}. 
\end{itemize} 
 
The intriguing  $\chi_{c0}(2P)$ was related to the  broad structures in $\bar D D$   invariant mass  in the same references \cite{Guo:2012tv,Olsen:2014maa}. The process $\gamma\gamma \to \bar D D$ from BaBar \cite{Aubert:2010ab} and Belle \cite{Uehara:2005qd} leads Guo\&Meissner   to\footnote{Here possible feed-down from $\gamma\gamma\to D^* \bar D$ followed by $D^*\to D\pi (\gamma)$ is ignored according to \cite{Abe:2007sya}. } 
\begin{equation} 
\label{guo_meissner}
\textrm{\cite{Guo:2012tv}}:\ \   m=3837.6\pm 11.5~\mathrm{MeV},\  \Gamma=221\pm 19 ~\mathrm{MeV},
\end{equation} 
while $e^+e^-\to J/\psi\, D\bar D$ from Belle \cite{Abe:2007sya} leads Olsen to 
\begin{equation}
\label{olsen}
\textrm{\cite{Olsen:2014maa}}:\ \ m=3878\pm 48~\mathrm{MeV}~,\  \Gamma=347 \genfrac{}{}{0pt}{}{+316}{-143} ~\mathrm{MeV}.
\end{equation}

Obviously the spectrum of scalar charmonia beyond the ground state presents an open question. Our aim is to shed some light on this issue by simulating $\bar DD$ scattering in $s$-wave on the lattice, and look for possible resonances in the extracted scattering matrix.  Preliminary results based on the same simulation and only one ensemble have been presented in Ref. \cite{Prelovsek:2013sxa}.
 
\begin{table}[t]
\begin{center}
\begin{tabular}{c|l|l}
			&	Ensemble (1)	&	Ensemble (2)            \cr
\hline			
$N_L^3\times N_T$ & $16^3\times32$&	$32^3\times64$	\cr
$N_f$ 			&	2		&	2+1				\cr
$a~$[fm] 		&	0.1239(13)	&	0.0907(13)		\cr
$L~$[fm] 		&	1.98(2)		&	2.90(4)				\cr
$m_\pi~$[MeV]		&	266(3)(3)	&	156(7)(2)		\cr
$L m_\pi$		&	2.68(3)	&	2.29(10)			\cr			
$\kappa_c~$(val)	&	0.12300		&	0.12686			\cr
\#configs 		&	279		&	196				\cr
\end{tabular}
\end{center}
\caption{\label{tab:ensembles} The gauge configurations of ensemble (1) are from \cite{Hasenfratz:2008fg,Hasenfratz:2008ce}. Those of ensemble (2) are provided by the PACS-CS collaboration \cite{Aoki:2008sm}.  $N_L$ and $N_T$ denote the number of lattice points in spatial and time directions, $N_f$  the number of dynamical flavors and $a$ the lattice spacing.}
\end{table}

\section{Lattice setup and charm-quark treatment}\label{sec_simulation}

The simulation is performed on two lattice ensembles with the parameters listed in   Table \ref{tab:ensembles}. Both ensembles have rather low $m_\pi L$ but this is not a serious issue for charmonia and $\bar D D$ scattering in this simulation, where pions do not enter explicitly. Further details about the ensembles and our implementation of charm quarks may be found in \cite{Hasenfratz:2008fg,Hasenfratz:2008ce,Lang:2011mn,Mohler:2012na} for ensemble (1) and in   \cite{Aoki:2008sm,Lang:2014yfa} for ensemble (2). 
  
To minimize heavy-quark discretization effects at finite lattice spacing the Fermilab method \cite{ElKhadra:1996mp,Oktay:2008ex} is used for the charm quarks. The corresponding dispersion relation \cite{Bernard:2010fr} for a meson $M$ containing charm quarks is 
\begin{equation}
E_M(p)=M_1+\frac{\mathbf{p}^2}{2M_2}-\frac{a^3W_4}{6}\sum_ip_i^4-\frac{(\mathbf{p}^2)^2}{8M_4^3}+ \dots\;,
\label{disp}
\end{equation}
where $\mathbf{p}=\frac{2\pi}{L}\mathbf{q}$ and $\mathbf{q}\in N^3$. 

On both ensembles the charm quark hopping parameter $\kappa_c$ is tuned  \cite{Mohler:2012na,Lang:2014yfa} using the spin-averaged charmonium mass  
\begin{equation}
\label{sa}
\bar m\equiv \frac{1}{4}(m_{\eta_c}+3 m_{J/\psi})~,\quad \bar m^{exp}=3.06859(17)~\mathrm{GeV}
\end{equation}
which is the relevant reference mass for our spectra of charmonium.  The $M_{1,2,4}$ for the spin-averaged charmonium were determined based on the lattice data from the lattice dispersion relation (\ref{disp}), setting $W_4$ to zero. Then $\kappa_c$ was fixed by tuning the kinetic mass $M_2$ to $\bar m^{exp}$. The corresponding values for the spin-averaged  $M_{1}$ are given in Table \ref{dispresults}.

\begin{table}[t]
\begin{center}
\begin{tabular}{c|ccc}
	meson 	& mass & 	Ensemble (1)	&	Ensemble (2)            \cr
\hline			
$D$ & $a M_1\equiv a m_D$		&	0.9792(11)	&	0.75317(83)	\cr
$D$ &  $a M_2$		                &	1.107(14)		&	0.840(22)		\cr
$D$ &  $a M_4$		                &	1.060(44)		&	0.95(15)		\cr 
\hline 
spin-aver. $\bar cc$ &   $a M_1\equiv a\bar m$	     &	1.52451(44) &   1.20444(15) 			 	\cr
\hline
$\bar D D$ vs. $\bar cc$ & $2m_D-\bar m$  &0.6910(36) & 0.6568(36)  \cr 
\end{tabular}
\end{center}
\caption{\label{dispresults}The parameters in the dispersion relation (\ref{disp}) for $D$ mesons and spin-averaged charmonium $\frac{1}{4}(m_{\eta_c}+3 m_{J/\psi})$. The last line is in GeV, others in lattice units.}
\end{table}

To investigate the $\bar DD$ scattering we  need the dispersion relation $E_D(p)$ for $D$ mesons, which is also given by Eq. (\ref{disp}) with parameters $M_{1,2,4}$  in Table \ref{dispresults}.  The common feature of spectra in the scalar and vector charmonium channel are two-particle states $\bar D D$   that have a discrete spectrum on the finite lattice. In the absence of interactions, $D(q)\bar D(-q)$ have energies according to (\ref{disp})
\begin{equation}
\label{ni}
E^{n.i.}_{D(q)\bar D(-q)}=2 E_D(\mathbf{q}\tfrac{2\pi}{L})~,\quad \mathbf{q}\in N^3~,
\end{equation}
which will be shifted due to the interaction.

Within the Fermilab approach, the rest masses have large discretization effects but mass differences are expected to be close to physical \cite{Kronfeld:2000ck} and can be compared to experiment. In order to compare the splitting $E^{lat}-\bar m^{lat}$  with  $E^{exp}-\bar m^{exp}$, we will sometimes plot
\begin{equation}
\label{e}
E=E^{lat}-\bar m^{lat}+\bar m^{exp}
\end{equation}
and compare  it with $E^{exp}$.   

An important quantity is the position of the $\bar DD$ threshold  with respect to our reference mass. The splitting $2m_D-\bar m$ for ensemble (2) is very close to the experimental  value  $2m_D^{exp}-\bar m^{exp}\simeq 0.666~$GeV,  while it is a bit larger for ensemble (1) due to the heavier pion mass and larger discretization effects (see Table \ref{dispresults}).  

Our charm quark treatment has been verified on ensemble (1) for  low-lying charmonia, $D$ meson resonances \cite{Mohler:2012na} and $D_s$ mesons \cite{Mohler:2013rwa,Lang:2014yfa}, where reasonable agreement with experiment was found. The  spectrum for $D_s$ mesons   and some other hadrons containing charm quarks were also determined on ensemble (2) \cite{Mohler:2013rwa,Lang:2014yfa} with even better agreement due to the lower pion mass and smaller discretization effects. 
  
\section{Analysis details}\label{sec_interpolators}

Interpolating fields $O$ are used to create and annihilate the physical system with $J^{PC}=1^{--}$ or $0^{++}$, isospin $I=0$ and total momentum zero. All quark fields in the interpolators are smeared according to the distillation  method $q\equiv \sum_{k=1}^{N_v}v^{(k)}v^{(k)\dagger}q_{point}$ \cite{Peardon:2009gh,Morningstar:2011ka}. We use $N_v=192$ eigenvectors of the lattice laplacian $v^{(k)}$ for ensemble (2) and $N_v=96$ or $64$ for ensemble (1). The distillation method is convenient  for calculating a variety of Wick contractions. The full distillation method  \cite{Peardon:2009gh}  is employed on ensemble (1) with a smaller volume and details of the implementation are given in \cite{Lang:2011mn,Mohler:2012na}. The stochastic version \cite{Morningstar:2011ka}  is used on ensemble (2) with larger volume and details of our implementation are provided in  \cite{Lang:2014yfa}.

\subsection{Vector channel}

$\bar D D$   in $p$-wave is the dominant two-meson contribution for $E\leq 4~$GeV, while $D_1 \bar D$ appears higher. Sixteen $\bar cc$ and two $\bar DD$ interpolating fields are used in the relevant irreducible representation $T_1^{--}$:
\begin{align} 
\label{O_vector}
O^{\bar cc}_{1-14}&=\bar c A_i c \;, \\
O_{15}^{\bar cc}&=R_{ijk}\bar c \gamma^j E^k c\ , \quad E_i\equiv Q_{ijk} \overleftarrow{\nabla_j}\overrightarrow{\nabla_k}  \;,\nonumber\\ 
O_{16}^{\bar cc}&=R_{ijk}\bar c \gamma_t\gamma^j E^k c \;,\nonumber \\
O_1^{DD}&=[\bar c \gamma_5 u(e_i)~\bar u\gamma_5 c(-e_i)\nonumber \\
&\ - \bar c \gamma_5 u(-e_i)~\bar u\gamma_5 c(e_i)]+\{u\to d\} \;,\nonumber\\ 
O_2^{DD}&=[\bar c \gamma_5 \gamma_t u(e_i)~\bar u\gamma_5 \gamma_t c(-e_i)\nonumber\\
&\ - \bar c \gamma_5 \gamma_t u(-e_i)~\bar u\gamma_5 \gamma_t c(e_i)]+ \{u\to d\}\;,\nonumber
\end{align}  
where $i$ denotes polarization, while  $Q_{ijk}$   and $R_{ijk}=R_{jik}$  are listed in \cite{Dudek:2007wv}. The $\bar cc$ interpolators $O_{1-14}^{\bar cc}$ for vector channel $T_1^{--}$ are listed in Table X of \cite{Mohler:2012na}.
The momentum is projected for each $D$ meson separately,  
\begin{equation}
\bar u\Gamma c(\mathbf{k})\equiv  \sum_{\mathbf{x}}e^{i2\pi \mathbf{k}\cdot\mathbf{x}/L} \bar u(\mathbf{x},t)\Gamma c(\mathbf{x},t)\;,
\end{equation}
so that the $O^{DD}$ couple to $p$-wave.   For ensemble (1) $N_v=64$ is used for $O_2^{DD}$, and  $N_v=96$ for  the remaining interpolators. 

The irreducible representation $T_1^{--}$ contains $J^{PC}=1^{--}$ states of interest, and also $\psi_3$ states with $J^{PC}=3^{--}$ coupling due to the broken rotational symmetry on the lattice. In the continuum limit, $O_{1-14}^{\bar cc}$ contain only $1^{--}$, while $O_{15,16}^{\bar cc}$ contain $1^{--}$ and $3^{--}$ \cite{Dudek:2007wv}, which will help us to identify the spin 3 admixture related to $\psi_3$.

\subsection{Scalar channel}

$\bar DD$ in $s$-wave and $J/\psi\, \omega$ are the dominant two-meson states in the energy region of interest $E\leq 4~$GeV. Seven $\bar cc$, four $\bar DD$, and two $J/\psi\, \omega$ interpolating fields  are used in the relevant irreducible representation $A_1^{++}$:
\begin{align} 
\label{O_scalar}
O^{cc}_{1-7}&=\bar c A c\;, \\
O^{DD}_1&=\bar c \gamma_5 u(0)~\bar u\gamma_5 c(0)+\ \{u\to d\} \;,\nonumber\\ 
O^{DD}_2&=\bar c \gamma_5 \gamma_t u(0)~\bar u\gamma_5 \gamma_t c(0)+\ \{u\to d\}\;,\nonumber\\ 
O^{DD}_3&=\sum_{e_k=\pm e_{x,y,z}}\bar c \gamma_5 u(e_k)~\bar u\gamma_5 c(-e_k)+\ \{u\to d\} \;,\nonumber\\ 
O^{DD}_4&=\sum_{|u_k|^2=2}\bar c \gamma_5 u(u_k)~\bar u\gamma_5 c(-u_k)+\ \{u\to d\} \;,\nonumber\\
O^{J/\psi\, \omega}_1&=  \sum_j \bar c \gamma_j c(0)~[\bar u\gamma_j u(0)+\ \{u\to d\}] \;,\nonumber\\  
O^{J/\psi\, \omega}_2&= \sum_j \bar c \gamma_j \gamma_t c(0)~[\bar u\gamma_j \gamma_t u(0)+\ \{u\to d\}]~. \nonumber
\end{align}  
$O_{1-7}^{\bar cc}$ are listed in Table X of \cite{Mohler:2012na}. The momenta are projected for each meson separately in  $O^{DD}$ and $O^{J/\psi\, \omega}$. For ensemble (1) $N_v=64$ is used for $O_{2,3}^{DD},~O^{J/\psi\, \omega}_2$, and  $N_v=96$ for  the remaining interpolators. 

The irreducible representation $A_1^{++}$ contains $J^{PC}=0^{++}$ states of interest, and in general also  states with $J\geq 4$, which appear at energies beyond our interest.  

The interpolator $O^{DD}_4$ is not used for ensemble (1) since $D(2)D(-2)$ appears above $4~$GeV. The  $O^{J/\psi\, \omega}$ are not used on ensemble (2) since the results from ensemble (1) indicate that $J/\psi\, \omega$ is almost decoupled from the rest of the system.\footnote{When the interpolators $O^{J/\psi\, \omega}$ are removed from the interpolator basis, the energies $E_n$ and overlaps $\langle O_k|n\rangle$ for the remaining eigenstates $n$ are practically unchanged for ensemble (1).}  

\begin{figure*}[htb!] 
\begin{center}
\includegraphics*[width=0.97\textwidth,clip]{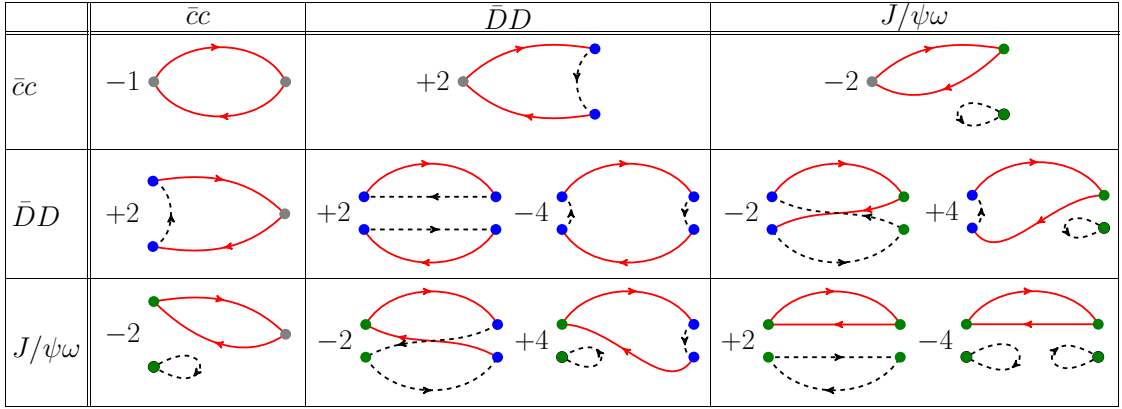} 
\caption{\label{fig:wick}  Wick contractions computed for the correlation matrix  (\ref{C}) with interpolators (\ref{O_vector},\ref{O_scalar}). We omit contractions where the charm quark annihilates.  A red solid line represents a $c$ quark, while the black dashed line represents a $u$ or $d$ quark.}
\end{center}
\end{figure*}

\subsection{Towards the spectrum}

The correlation matrix 
\begin{align}
\label{C}
C_{jk}(t)&= \langle \Omega| O_j (t^\prime +t) O_k^\dagger (t^\prime)|\Omega \rangle
=\sum_nZ_j^nZ_k^{n*}~e^{-E_n t}~
\end{align}
contains the information on energies  $E_n$ and the overlaps $Z_j^n\equiv \langle \Omega|{\cal O}_j|n\rangle$.  
We evaluate  all Wick contractions for $O\simeq \bar cc,~(\bar qc)(\bar cq), ~(\bar cc)(\bar qq)$ (\ref{O_vector},\ref{O_scalar}) shown in Fig.  \ref{fig:wick}. We omit Wick contractions where charm quark annihilates as in almost all previous lattice simulation of charmonia; these induce mixing with $I=0$ decay channels containing only light quarks $u,d,s$, they are Okubo-Zweig-Iizuka suppressed and present a challenge for current lattice simulations. It is noteworthy that these decays might be important to clarify the experiment puzzle with regard to non-$\bar{D}D$ hadronic decays \cite{Rosner:2004wy,Wang:2004kf}. 

The energies and overlaps are extracted from the correlation matrix using  the generalized
eigenvalue method \cite{Michael:1985ne,Luscher:1985dn,Luscher:1990ck,Blossier:2009kd}
\begin{align} 
 C(t)u^{(n)}(t)&=\lambda^{(n)}(t)C(t_0)u^{(n)}(t)~,
\end{align}
where $\lambda^{(n)}(t)\propto  e^{-E_n t}$ at large $t$.  Correlated two or one-exponential fits to $\lambda^{(n)}(t)$ are used and $t_0=2,~3$.  The errors-bars correspond to statistical errors obtained using single-elimination jack-knife.

\begin{figure*}[htb] 
\begin{center}
\includegraphics*[width=0.70\textwidth,clip]{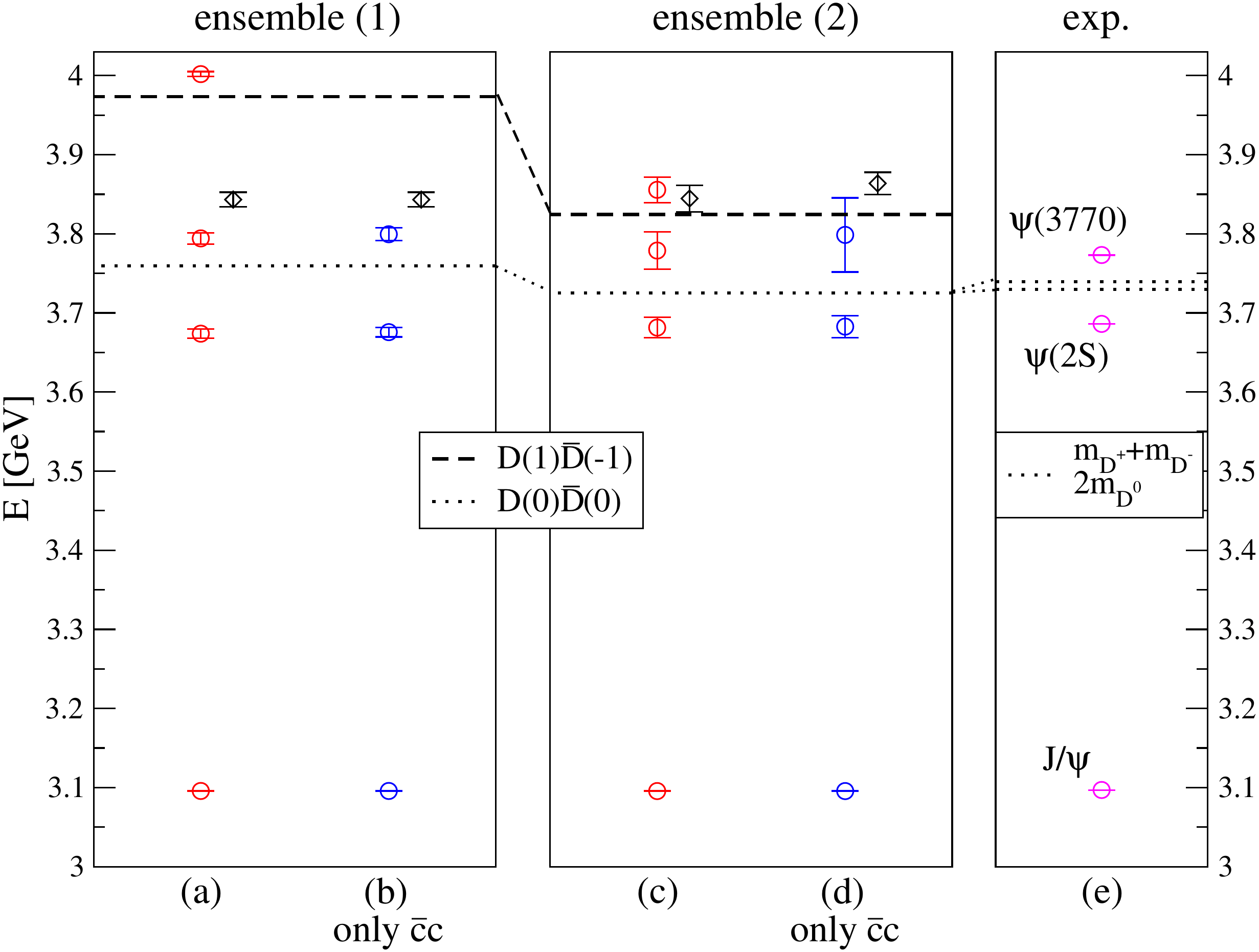} 
\caption{\label{fig:e_vector} The  energies $E$ (see Eq. (\ref{e})) in the vector channel  on both ensembles, together with the experimental masses. The circles represent $J^{PC}=1^{--}$ states, while the diamond represents a $3^{--}$ admixture present in the irreducible representation $T_1^{--}$ and related to the $\psi_3$.  The dashed lines show the non-interacting energy of $D(1)\bar D(-1)$ (\ref{ni}), and  the dotted line  represents the threshold $2m_D$. The $D(0)\bar D(0)$ state does not appear for $p$-wave.  Interpolators used in (a,c) are given in Table \ref{tab:e_vector}, while (b,d) utilize just $O^{\bar cc}$ from the same sets. }
\end{center}
\end{figure*}  

\section{Results for the vector  channel}\label{sec:results_vector}
 
\subsection{Discrete spectrum}
 
\begin{table}[t]
\begin{center}
\footnotesize
\begin{tabular}{ccccccccccc}
n  & fit & fit & $\tfrac{\chi^2}{d.o.f.}$ & $E^{lat}a$ & $E~$[GeV]& $(ap)^2$ & $(ap)^3 \cot(\delta)$ & $\frac{(ap)^3 \cot(\delta)}{\sqrt{s}}$& $\delta[^\circ]$\cr
 & range & type & & & (\ref{e})& & & \cr
\hline 
 &  Ens. &(1)      &        &           &               &               &              &                &                &                 \cr
1 &$3$-$14$&$2e^c$&$8.57/8$ &$1.54153(43)$ &$3.09572(34)$ &      /       &      /        &      /        &      /         \cr
2 &$3$-$14$&$2e^c$&$16.62/8$&$1.9045(38)$  &$3.6738(58)$  &$-0.0588(47)$ &$0.0137(18)$   &$0.00717(95)$  &$-109.1(6.4)\I$ \cr
3 &$3$-$13$&$2e^c$&$5.18/7$ &$1.9801(46)$  &$3.7941(71)$  &$0.02413(57)$ &$-0.00599(34)$ &$-0.00303(17)$ &$148.0(7.7)$    \cr
4 &$3$-$13$&$2e^c$&$5.09/7$ &$2.0109(60)$  &$3.8433(93)$  &      /       &      /        &      /        &      /         \cr
5 &$3$-$13$&$2e^c$&$8.49/7$ &$2.1105(21)$  &$4.0019(32)$  &$0.1755(33)$  &$-0.144(23)$   &$-0.068(11)$   &$153.0(4.4)$    \cr
 & Ens. &(2)      &        &           &               &               &               &                &                &                 \cr
1 &$3$-$29$&$2e^c$&$3.15/23$&$1.21683(16)$&$3.09557(18)$&      /       &      /       &       /       &      /         \cr
2 &$3$-$11$&$2e^c$&$3.44/5$ &$1.4862(60)$ &$3.682(13)$  &$-0.0169(50)$ &$0.0021(10)$  & $0.00143(68)$ &$120(25)\I$     \cr
3 &$3$-$11$&$2e^c$&$4.36/5$ &$1.531(11)$  &$3.779(24)$  &$0.0207(93)$  &$0.00056(255)$& $0.00037(167)$&$79(40)$        \cr
4 &$3$-$11$&$2e^c$&$4.92/5$ &$1.5611(78)$ &$3.845(17)$  &      /       &      /       &       /       &      /         \cr
5 &$3$-$11$&$2e^c$&$4.78/5$ &$1.5661(75)$ &$3.856(16)$  &$0.0509(65)$  &$-0.0054(76)$ & $-0.0034(49)$ & $115(35)$     \cr	
\end{tabular}
\normalsize
\end{center}
\caption{\label{tab:e_vector}  Discrete lattice spectrum from charmonium in the irreducible representation $T_1^{--}$ which contains $J^{PC}=1^{--}$, $3^{--}$ and higher $J$ states. The $p$ and $\delta$ correspond to $\bar DD$ scattering in $p$-wave. Subset  $O^{\bar cc}_{1-6,8,9,11,12,15},O^{DD}_{17,18}$ from the interpolators in Eq. (\ref{O_vector}) is used for ensemble (1) and $O^{\bar cc}_{1,3-5,9-11,13,15},O^{DD}_{17}$ for ensemble (2).  $t_0=2$ is used for all data points. }
\end{table}

The energy levels  in the vector channel are shown in Fig. \ref{fig:e_vector}a and \ref{fig:e_vector}c together with the experimental masses. The full set of operators gave noisier signals than suitable subsets, and  the chosen subsets are listed in Table \ref{tab:e_vector}.   The circles denote the energy levels that are related to $J^{PC}=1^{--}$  states $J/\psi,~ \psi(2S),~\psi(3770),~D(1)D(-1)$ (from bottom to top), while $D(0)\bar D(0)$ does not appear for $p$-wave. The diamond indicates a level related to the $J^{PC}=3^{--}$ state $\psi_3$, that is present in representation $T_1^{--}$ due to the broken rotational symmetry on the lattice.  

\begin{figure*}[htb]
\begin{center}
\includegraphics[width=0.98\textwidth,clip]{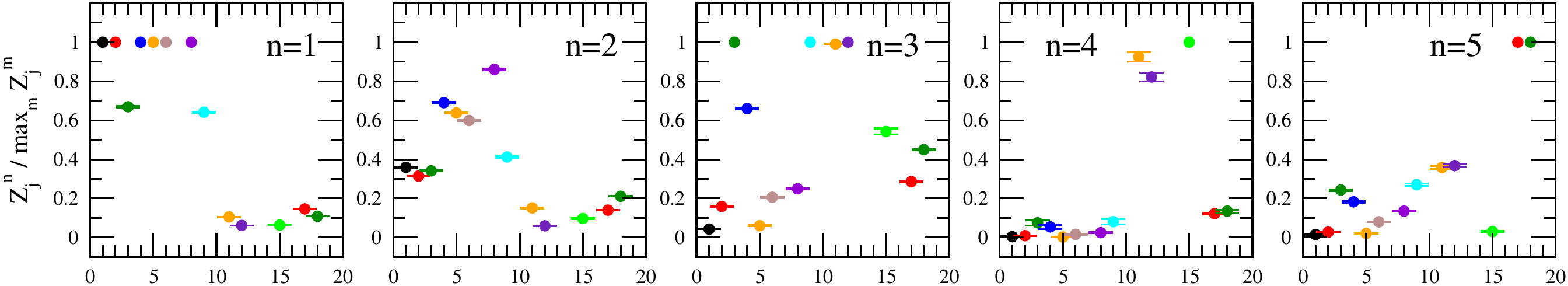}\\
\includegraphics[width=0.98\textwidth,clip]{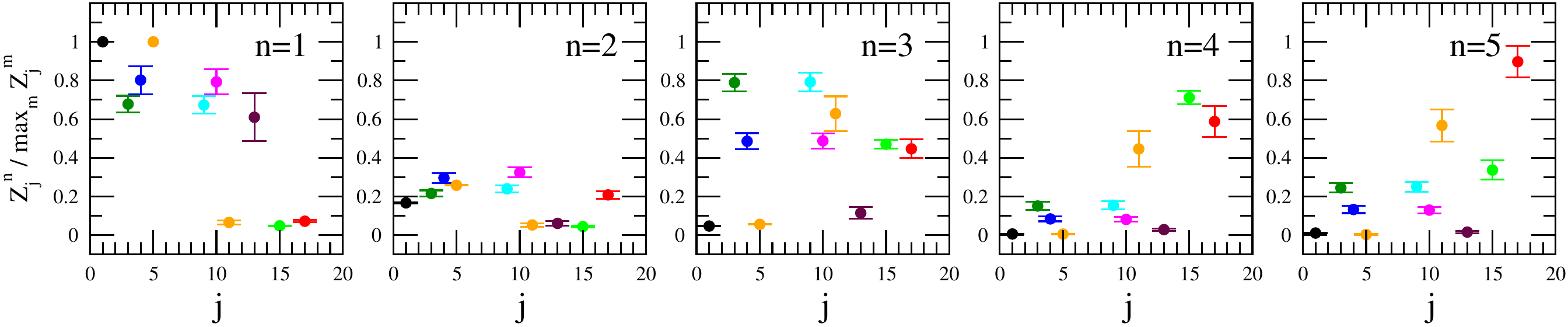}
\end{center}
\caption{  The overlaps $Z_j^n=\langle \Omega|{\cal O}_j|n\rangle$  for the vector channel show the matrix elements of interpolators ${\cal O}_j$ between the vacuum $\langle \Omega|$ and the  eigenstate $|n\rangle$ on the lattice. We present the overlap ratios $Z_j^n/\mathrm{max}_m Z_j^m$ on ensemble (1) (top) and on ensemble (2) (bottom). The denominator  is  the maximal  $|Z_j^m|$ at given operator number $j$. These ratios are  independent on the normalization of the interpolators $O_j$.  Levels $n=1,..,5$  are ordered  from lowest to highest $E_n$ in Figs. \ref{fig:e_vector}a and  \ref{fig:e_vector}c for both ensembles, respectively.  The order of interpolators $j$ on the abscissa (listed in caption of Fig. \ref{tab:e_vector}) is the same as in the list (\ref{O_vector}).}\label{fig:z_vector} 
\end{figure*}  

\begin{table}[t]
\begin{center}
\begin{tabular}{c|ccc}
 ground st.   & 	Ensemble (1)	&	Ensemble (2)            \cr
\hline			
 $E^{lat}a$   &     $2.0124(38)$     &     $1.559(51)$       \cr 
\end{tabular}
\end{center}
\caption{\label{tab:A2} The energy of $\psi_3$ with $J^{PC}=3^{--}$ from the ground state in the $A_2^{--}$ irreducible representation of the $O_h$ point group. }
\end{table}

The highest state ($n=5$) has largest overlap with $O^{DD}$ and disappears when these interpolators are excluded from the basis, as shown in Fig.  \ref{fig:e_vector}b and \ref{fig:e_vector}d. Each energy level in addition to $D(1)\bar D(-1)$ indicates the presence of a bound state or a resonance.  Good resemblance with the experimental spectrum is indeed confirmed in Fig. \ref{fig:e_vector}.  The $J/\psi$ is significantly below threshold and no effect from threshold is expected. The $\psi(2S)$ is situated $\simeq 42~$MeV below threshold in  experiment, and the corresponding finite volume energy on the lattice does not depend (within uncertainties) on whether $\bar D D$  interpolators  are used or not (see Fig. \ref{fig:e_vector}).  The appearance of levels $n=3$ and $4$ is related to the $\psi(3770)$ resonance and to the spin 3 admixture and the corresponding $\psi_3$ resonance. Level $n=4$ is related to $\psi_3$ due to smaller overlaps $\langle O^{\bar cc}_{1-14}|n=4\rangle$. This is based on the fact  that $O^{\bar cc}_{1-14}$ couple in the continuum limit only to $1^{--}$ (which is responsible for small $\langle O^{\bar cc}_{1-14}|\psi_3\rangle$ at finite $a$), while $O^{\bar cc}_{15,16}$ couple to  $1^{--}$ and $3^{--}$. Further support is given by the near-degeneracy with the energies from the irreducible representation $A_2^{--}$ where a $3^{--}$ state comes as the ground state (see Table \ref{tab:A2}). For the $\psi(3770)$ the avoided-level crossing scenario suggests $E_3$ in the energy region $m\pm \Gamma$, which is reasonably satisfied by comparing to experiment. In order to really determine the resonance mass and width for $\psi(3770)$ one needs to consider the phase shifts for $\bar D D$ scattering in $p$-wave.

\subsection{\texorpdfstring{$\bar D D$}{DD} scattering in  \texorpdfstring{$p$-wave}{pwave}}

We assume that $\bar D D$ scattering in $p$-wave near the resonance $\psi(3770)$ is elastic, which is a good approximation since $Br[\psi(3770)\to D\bar D] =93 \pm 9 \%$, while the remaining part goes mainly  to  light hadrons and charmonium states (i.e. $J/\psi \pi\pi$, $J/\psi \eta$\ldots). In the elastic  case, the scattering phase shift $\delta$ is given by L\"uscher's relation  \cite{Luscher:1990ux,Luscher:1991cf}
\begin{equation}
\label{luscher}
p\cot\delta(p)=\frac{2Z_{00}(1;(\tfrac{pL}{2\pi})^2)}{L\sqrt{\pi}}~,
\end{equation}
which applies for the total momentum zero employed in our case. The momentum $p$ of $D$ mesons is extracted from the measured energy levels $E_n^{lat}=2E_D(p)$ using the dispersion  relation (\ref{disp}). The resulting momenta and phase shifts for all eigenstates except for the spin 3 admixture and for the finite volume state related to $J/\psi$ are collected in Table \ref{tab:e_vector}. The large absolute value of $p^3\cot\delta$ corresponds to feeble scattering, while small $p^3\cot\delta$ is related to significant scattering.

We fit our data in two ways:

\begin{figure}[bh]
\begin{center}
\includegraphics[width=0.49\textwidth,clip]{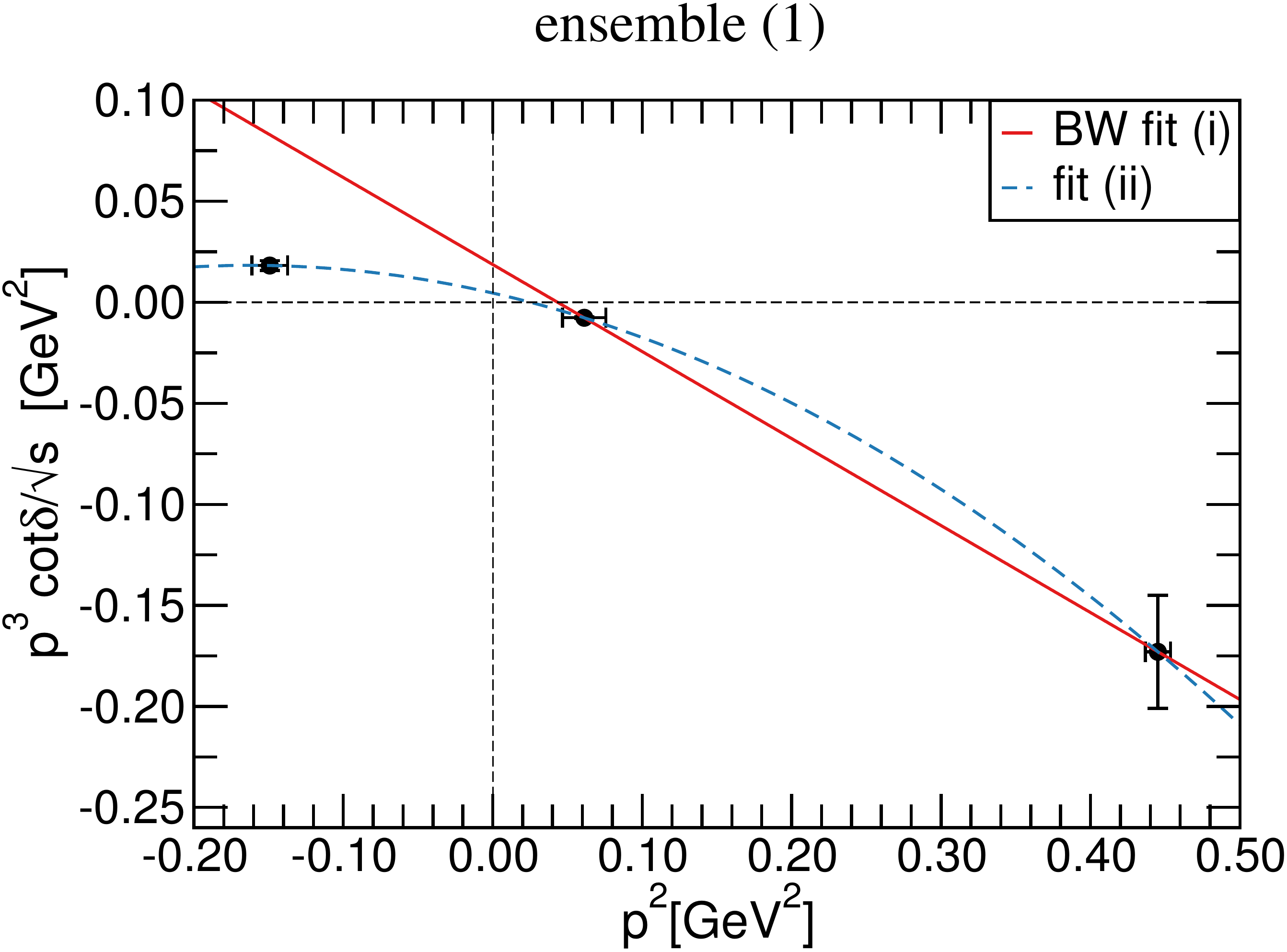}
\includegraphics[width=0.49\textwidth,clip]{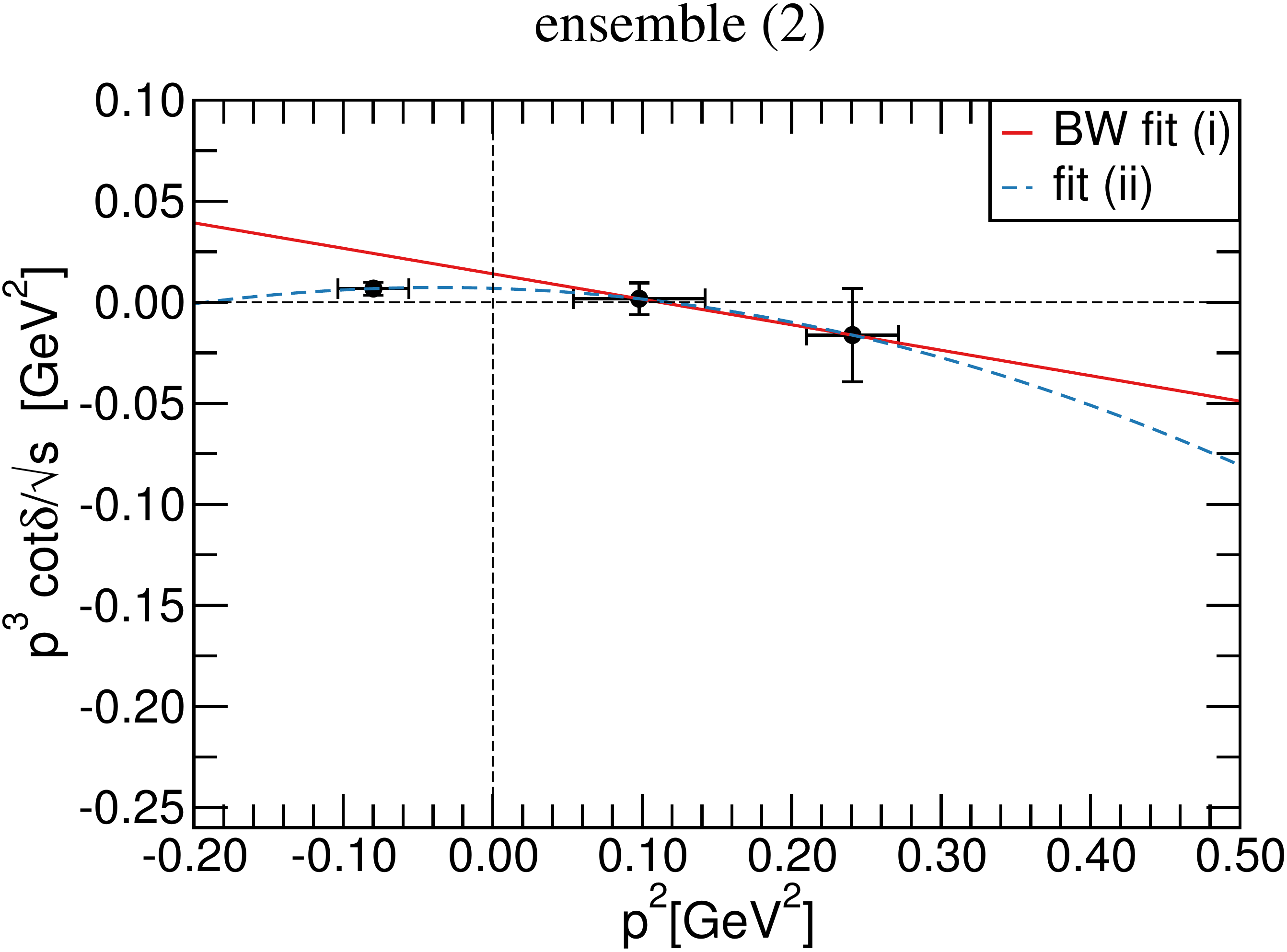}
 \end{center}
\caption{$p^3 \cot\delta/\sqrt{s}$ versus $p^2$ for $\bar D D$  scattering in $p$-wave in the region of the $\psi(2S)$ bound state and the $\psi(3770)$ resonance. The $p$ denotes  the momentum of $D$ meson. We show the Breit-Wigner fit (i) and the extended fit (ii), which aims to capture also the behavior around $\psi(2S)$.}\label{fig:phase_vector} 
\end{figure}

\begin{table}[tbh]
\begin{center}
\begin{tabular}{c|cc|cc|c}
 & 	\multicolumn{2}{c|}{Ensemble (1)}    &	\multicolumn{2}{c|}{Ensemble (2)}            & exp \cr
 &                 fit (i)   &   fit (ii)   & fit (i)     & fit (ii)       & $D^+ D^-$/$D^0\bar D^0$\cr
\hline
$\psi(3770)$  & & &&&\cr
$p_R~$[GeV]   & $0.208(31)(3)$  & $0.159(35)(2)$  & $0.334(155)(5)$ & $0.343(184)(5)$    & $0.26/0.29$\cr 		
$m_R~$[GeV]   & $3.784(7)(10)$  & $3.774(6)(10)$  & $3.786(56)(10)$ & $3.789(68)(10)$  & $3.77315(33)$    \cr  
$g$ (no unit) & $13.2(1.2)$     & $19.7(1.4)$     & $24(19)$        & $28(21)$         & $18.7(1.4)$     \cr 
\hline
$ \psi(2S)$   &                 &                 &                 &                  &\cr
$|p_B|~$[GeV] &                 & $0.380(17)(6)$  &                 & $0.280(43)(4)$  & $0.31/0.28$\cr 			
$m_B~$[GeV]   &                 & $3.676(6)(9)$   &                 & $3.682(13)(9)$  & $3.686109\genfrac{}{}{0pt}{}{+12}{-14}$\cr 
\end{tabular}
\end{center}
\caption{\label{tab:res_vector}  Parameters of the resonance $\psi(3770)$ and bound state $\psi(2S)$  from fits (i) (\ref{fit_1_vector}) and (ii) (\ref{fit_2_vector}). The $\psi(3770)\to D\bar D$ width  $\Gamma=g^2p^3/(6\pi s)$ is parametrized in terms of the coupling $g$ and compared the value of the coupling derived from experiment \cite{pdg14}. The $p_{R}$ denotes $D$-meson momenta at the peak of the resonance and $|p_B|$ the binding momentum. The first errors are statistical and the second errors (where present) are from the scale setting uncertainty. The experimental data and errors are based on PDG values. Errors on experimental $p_{R/B}$ are suppressed as they are very small. }
\end{table}

\noindent{\bf (i) A resonance $\mathbf {\psi(3770)}$:}
The scattering matrix in the vicinity of a resonance  has a Breit-Wigner form 
\begin{equation}
\label{amplitude}
T_l(s)=\frac{\sqrt{s}\,\Gamma(s)}{m^2_{R}-s-i \sqrt{s}~
\Gamma(s)} =\frac{1}{\cot \delta_l(s)-i}~.
\end{equation}
The width 
\begin{equation}
\label{gam_scalar} 
\Gamma(s)=\frac{g^2}{6\pi}\frac{p^{3}}{s}
\end{equation}
is parametrized in terms of the phase space for $p$-wave decay and the $\psi(3770)\to D\bar D$ coupling $g$. It is expected that the leading dependence of $\Gamma$ on $m_{u/d}$ is captured by phase space. Equations (\ref{amplitude}, \ref{gam_scalar}) lead to  $p^{3}\cot \delta_1(s)/\sqrt{s}=(6\pi/g^2)(m_{R}^2-s)$ and then expressing  $s=2(m_D^2+p^2)^{1/2}$ and $m_R=2(m_D^2+p_R^2)^{1/2}$ to 
\begin{equation}
\label{fit_1_vector}
\frac{p^{3}\cot\delta(s) }{\sqrt{s}}=\frac{6\pi}{g^2}4(p_{R}^2-p^2)~
\end{equation}
where $p_R$ is the $D$ meson momentum at the resonance peak. The values of $g$ and $p_R$  follow from the linear fit (\ref{fit_1_vector}) through the energy levels $n=3,5$  in the vicinity of the resonance, where the  Breit-Wigner form  applies (level 4 is omitted since it is attributed to $\psi_3$ as discussed above). The fit is shown in Fig. \ref{fig:phase_vector}, while the resulting resonance parameters are given in   Table \ref{tab:res_vector}.  The resonance mass $m_R$ corresponding to the $p_R$ on the lattice is given by inserting the Fermilab dispersion relation (\ref{disp}) in (\ref{e})  
\begin{equation}
\label{mR}
m_{R/B}=2E_D(p_{R/B})-\bar m^{lat}+\bar m^{exp}
\end{equation}
and will be used for resonances or bound states throughout this work.\\
 
\noindent{ \bf (ii) A resonance $\mathbf {\psi(3770)}$ and  a bound state $\mathbf {\psi(2S)}$:}
In addition to the  Breit-Wigner form (\ref{fit_1_vector}), which is linear in $p^2$, we make use also of the  square form in $p^2$
\begin{equation}
\label{fit_2_vector}
\frac{p^{3}}{\sqrt{s}}\cot\delta_1(s)=A+Bp^2+Cp^4
\end{equation}
which in general has a longer range of applicability. It aims to capture also the $\bar DD$ scattering in the vicinity of $\psi(2S)$: there the (imaginary) phase shift  in Table \ref{tab:e_vector} nearly satisfies the condition for the bound state $\cot\delta\simeq i$ on the physical Riemann sheet $p_B=i|p_B|$, leading to $p^3\cot\delta\simeq |p_B|^3$.  The fit (\ref{fit_2_vector}) through levels $n=2,3,5$ in Fig. \ref{fig:phase_vector} assumes that the $\psi(2S)$ state still affects the $\bar D D$   scattering. It renders $(A,B,C)\simeq (0.0046(19)~\mathrm{GeV}^2,-0.168(27),-0.52(13)/\mathrm{GeV}^{2})$, $(0.0069(88)~\mathrm{GeV}^2,-0.023(80),-0.30(68)/\mathrm{GeV}^{2})$ on ensemble (1) and ensemble (2) respectively. The zero of $p^{3}\cot\delta_1/\sqrt{s}$, and the derivative at this zero,   lead to the parameters of $\psi(3770)$ resonance in Table \ref{tab:res_vector}. This model also leads to a bound state $\psi(2S)$ at $p_B=i|p_B|$ where the scattering amplitude $T$ (\ref{amplitude}) has a pole and $\cot \delta(p_B)=i$. The bound state mass $m_B$ in  Table \ref{tab:res_vector} is indeed close to experimentally measured $\psi(2S)$.
  
On ensemble (2) the results both from fit~(i) and fit~(ii) are compatible with the experimental data\footnote{Since we work in the isospin-symmetric limit we measure the sum of the neutral and charged decay modes; therefore we compare to the experimental value  $g_{exp}^2=g^2_{D^0\bar D^0}+g^2_{D^+ D^-}$ obtained from $\Gamma[\psi(3770)\to D_0 \bar D_0]=g^2_{D^0\bar D^0} p^3/(6\pi s)$ and  $\Gamma[\psi(3770)\to D^+ D^-]=g^2_{D^+ D^-} p^3/(6\pi s)$ \cite{pdg14}. Notice also that averaging the results from recent experiment resonance mass determinations for the $\psi(3770)$ leads to a value of $m_R^{exp}=3778.1(1.2)$, much larger than the fit by the PDG (which relies on an experiment neglecting interference with non-resonant background) and consistent with the most recent results in \cite{Anashin:2011kq}.} within large statistical uncertainties (see Table \ref{tab:res_vector}). Note that the higher-lying $\psi(4040)$ resonance  does not influence the results (for this ensemble), since it lies  significantly higher than the relevant energy levels.

On ensemble (1) the results for $\psi(3770)$ from fit~(i) give a smaller resonance momentum $p_R$ than in experiment, which we attribute to the unphysical threshold on ensemble (1) at $m_{\pi} \simeq 266~$MeV and the finite lattice spacing. The resonance mass $m_R$ calculated as in Eq. \ref{e} compares favorably. The coupling constant from fit~(i) is to small compared to experiment which is likely related to to the closeness of the $\psi(4040)$ resonance neglected in the analysis. The assumption that the resonance $\psi(4040)$ does not affect the energy level related to $D(1)\bar D(-1)$ is probably not justified on ensemble (1), where energy level lies higher (and closer to $\psi(4040)$) than on ensemble (2).  Roughly estimating the effect by comparing the one-resonance and two-resonance scenarios, estimating $g$ and $p_R$ for $\psi(3770)$ and $\psi(4040)$ from available experimental data \cite{Aubert:2009aq}, the coupling we observe is consistent with this interpretation \footnote{The maximal effect  of $\psi(4040)$ is estimated by assuming that $\psi(4040)$ width is saturated by $D\bar D$ (instead of $D\bar D$, $D\bar D^*$,  $D^*\bar D^*$ and other modes).}. Given the possibly large influence from the $\psi(4040)$ we can not conclude that fit (ii) is better than fit (i) on this ensemble.

\begin{figure}[htb] 
\begin{center}
\includegraphics*[width=0.70\textwidth,clip]{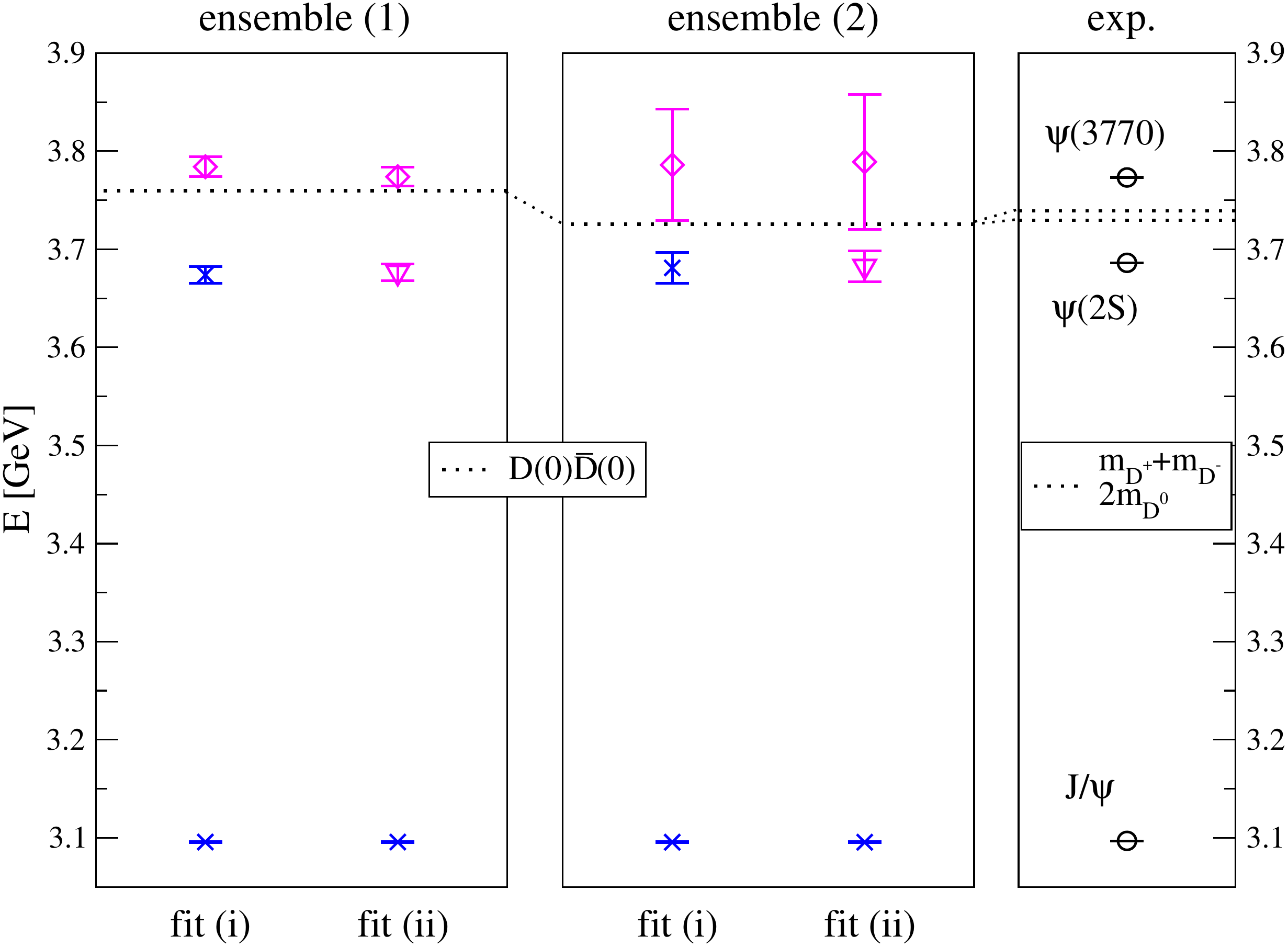} 
\caption{\label{fig:summary_vector}   The  comparison of the final $1^{--}$ spectrum to  the experiment. The magenta diamond denotes $\psi(3770)$ resonance mass from the Breit-Wigner fit (i) or extended fit (ii), given in Eqs. (\ref{fit_1_vector}) and (\ref{fit_2_vector}), respectively.  The magenta triangle denotes $\psi(2S)$ obtained as a pole in    $D\bar D$ channel. The blue triangles denote masses of $J/\psi$ and $\psi(2S)$ extracted as energy levels in the finite box. The statistical and scale setting errors have been summed in quadrature. }
\end{center}
\end{figure}

The  resulting $1^{--}$ spectrum   is summarized and  compared to experiment in Fig. \ref{fig:summary_vector}. 

\begin{figure*}[htb] 
\begin{center}
\includegraphics*[width=0.70\textwidth,clip]{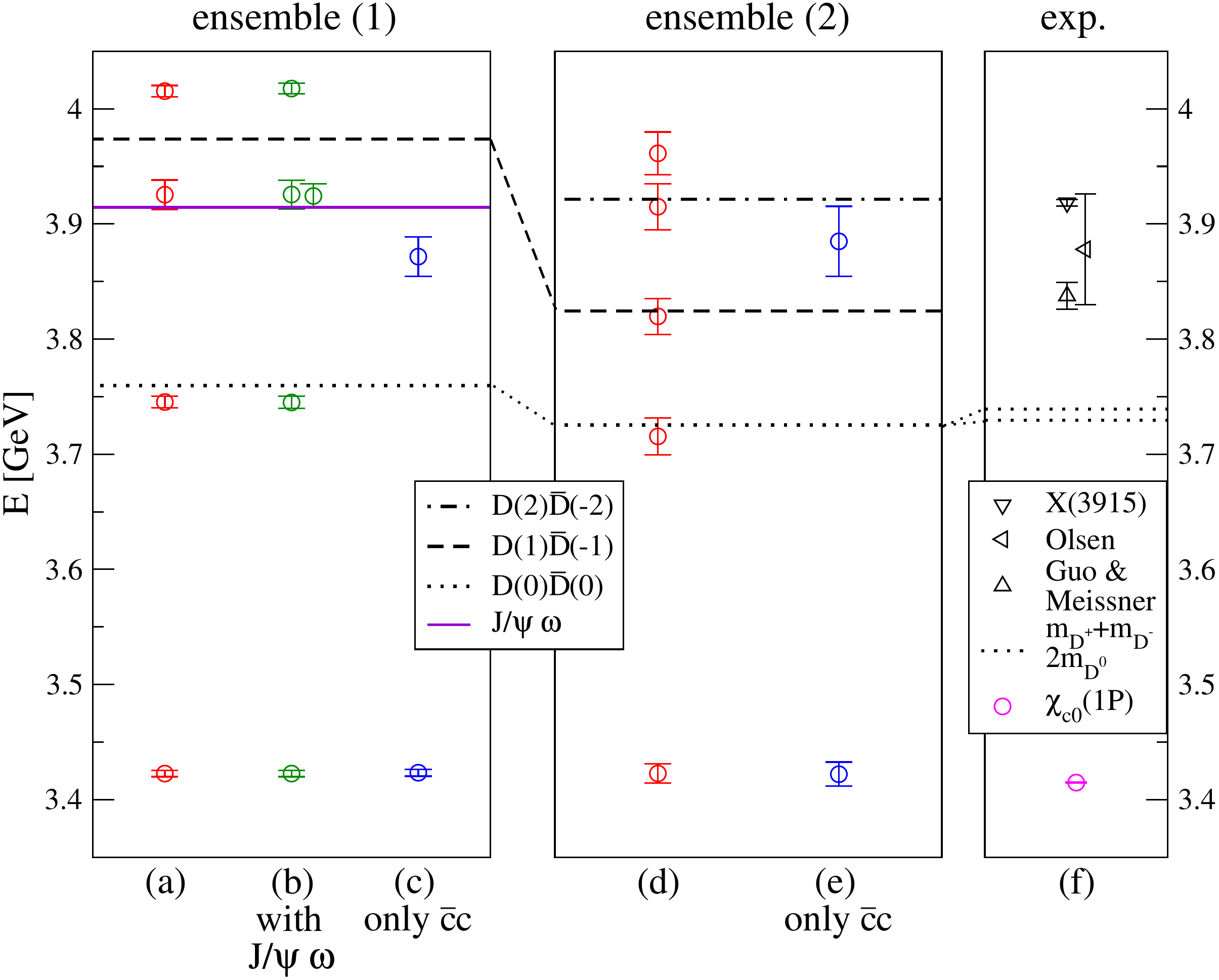} 
\caption{\label{fig:e_scalar}   The  energies $E$ (see Eq. \ref{e}) in the scalar channel  on both ensembles.   The only well established experimental state $\chi_{c0}(1P)$ is shown by the (magenta) circle. Triangles show three intriguing candidates for $\chi_{c0}(2P)$, that are not universally accepted: $X(3915)$ and  the broad resonances (\ref{guo_meissner},\ref{olsen}) suggested in \cite{Guo:2012tv,Olsen:2014maa}. The dashed lines shown energies of non-interacting $D(q)\bar D(-q)$ with $q=0,1,2$ (\ref{ni}), while dot-dashed line represents $m_{J/\psi}+m_\omega$.  Interpolators used in (a,d) are given in Table \ref{tab:e_scalar}, (b) uses $O^{J/\psi\, \omega}_{1,2}$ in addition, while (c,e) are based only on $O^{\bar cc}_{1,3,5}$. }\end{center}
\end{figure*}

\section{Results for the scalar channel}\label{sec:results_scalar}

\subsection{Discrete spectrum}

The energy levels  in the scalar channel are shown in Figs. \ref{fig:e_scalar}.  The only experimentally well established state is $\chi_{c0}(1P)$. The triangles represent the intriguing experimental candidates for $\chi_{c0}(2P)$, none of which is commonly accepted (see Section \ref{sec:open_questions}).  

 The spectrum from a lattice simulation consists both of energy levels that have large overlap with $\bar q q$ operators as well as energy levels with dominant overlap to $\bar D D$ operators. The latter appear near their non-interacting energies $E^{n.i.}_{DD}$  of Eq.  (\ref{ni}), which are denoted by dashed lines in Figs. \ref{fig:e_scalar}a,d. On ensemble (1) levels $n=2,4$ appear near the non-interacting $D(0)\bar D(0)$ and $D(1)\bar D(-1)$ (cf. Fig. \ref{fig:e_scalar}a). Levels $n=2,3,4$ on ensemble (2) have dominant overlap to $\bar D D$ scattering operators and are close to non-interacting $D(0)\bar D(0)$, $D(1)\bar D(-1)$ and $D(2)\bar D(-2)$ energies (cf. Fig. \ref{fig:e_scalar}d).
   
In the elastic case each energy level in addition to the number of expected $D(q)\bar D(-q)$ scattering levels is related to the presence of a bound state or a resonance. There are two such states, that cannot be attributed to $D(q)\bar D(-q)$ for both ensembles. The ground state is related to $\chi_{c0}(1P)$ and is close to its experimental mass. The second of these two levels appears above threshold and corresponds to $n=3$ for ensemble (1) and $n=5$ for ensemble (2), as shown in Figs.  \ref{fig:e_scalar}a and  \ref{fig:e_scalar}d. The avoided level crossing scenario suggests that an additional level appears somewhere in the range $E\simeq m\pm \Gamma$, which suggests the existence of a resonance roughly at
 \begin{equation}
\label{naive_scalar}
  m  \simeq 3.9-4.0 ~\mathrm{GeV}~\quad (\mathrm{naive\ estimate\ from}\ E_n)\;.
\end{equation}
This is close to  the first excitation obtained using just $O^{\bar cc}$ interpolators in Figs.   \ref{fig:e_scalar}c and  \ref{fig:e_scalar}e. Such a basis gives a rough estimate of resonance masses but is not well suited to capture two-particle states or resonances and bound states close to threshold \cite{Lang:2011mn,Dudek:2012xn,Mohler:2012na}.  
 
The  spectrum including $J/\psi\, \omega$ interpolating fields is shown in Fig. \ref{fig:e_scalar}b for ensemble (1). An energy level related to $J/\psi(0)\,\omega(0)$ appears at roughly $m_{J/\psi }+m_\omega$ while the energies of all the other levels remain unaffected with respect to Fig. \ref{fig:e_scalar}a. We have verified also that the overlaps for the remaining levels  are not affected if $O^{J/\psi\, \omega}$ are in the basis or not. This indicates that the $J/\psi\, \omega$ channel is decoupled from $\bar D D$   channel to a good approximation. 
   
\subsection{\texorpdfstring{$\bar D D$}{DD2} scattering in \texorpdfstring{$s$-wave}{swave} }

\begin{table}[t]
\begin{center}
\footnotesize
\begin{tabular}{cccccccccccc}
n & fit & fit & $\tfrac{\chi^2}{d.o.f.}$ & $E^{lat}a$ & $E~$[GeV] & $(ap)^2$ & $(ap) \cot(\delta)$ & $\frac{(ap) \cot(\delta)}{\sqrt{s}}$ & $\delta [^{\circ}]$ \cr
& range  & type &  & & (\ref{e}) &   & &  \cr
\hline 
&  Ens.   & (1)       &        &            &               &               &               &               &                &   \cr
1 & $6$-$15$ & $2e^c$ & $9.50/6$ & $1.7468(19)$  & $3.4226(27)$ & $-0.2226(49)$ & $-0.4716(52)$ & $-0.2700(31)$  & $-240.9(2.8)\I$ \cr
2 & $6$-$15$ & $2e^c$ & $5.40/6$ & $1.9494(33)$  & $3.7453(52)$ & $-0.0099(32)$ & $0.11(11)$    & $0.058(55)$    & $4(190)+84(306)\I$ \cr
3 & $3$-$12$ & $2e^c$ & $2.73/6$ & $2.0625(81)$  & $3.925(13)$  & $0.1185(98)$  & $0.39(17)$    & $0.191(79)$    & $41(11)$        \cr
4 & $3$-$12$ & $2e^c$ & $6.39/6$ & $2.1190(31)$  & $4.0154(49)$ & $0.1857(48)$  & $-0.50(11)$   & $-0.236(52)$   & $139.2(6.5)$    \cr
 & Ens.    &  (2)      &        &            &               &               &               &               &                &  \cr
1 & $4$-$15$ & $2e^c$ & $7.84/8$ & $1.3672(38)$  & $3.4227(83)$ & $-0.1136(60)$ & $-0.3371(89)$ & $-0.2466(68)$  & $-344.9(8.8)\I$ \cr
2 & $4$-$15$ & $2e^c$ & $3.74/8$ & $1.5018(74)$  & $3.715(16)$  & $-0.0038(61)$ & $-0.004(185)$ & $-0.003(123)$  & $-4(201)\I$     \cr
3 & $3$-$12$ & $2e^c$ & $3.50/6$ & $1.5497(71)$  & $3.820(16)$  & $0.0367(61)$  & $1.2(7.5)$    & $0.8(4.8)  $   & $8.7(29.2)$     \cr
4 & $6$-$12$ & $1e^c$ & $0.89/5$ & $1.5934(92)$  & $3.915(20)$  & $0.0745(80)$  & $1.7(10.8)$   & $1.1(6.8)$     & $9.1(28.3)$     \cr
5 & $6$-$12$ & $1e^c$ & $1.68/5$ & $1.6148(85)$  & $3.961(19)$  & $0.0932(75)$  & $-0.21(21)$   & $-0.13(13)$    & $124(28)$       \cr
\end{tabular}
\normalsize
\end{center}
\caption{\label{tab:e_scalar}  Discrete lattice spectrum  for the scalar channel. The $p$ and $\delta$ correspond to $\bar DD$ scattering in $s$-wave.  
Subset  $O^{\bar cc}_{1,3,5},O^{DD}_{1-3}$ from interpolators in Eq. (\ref{O_vector}) is used for ensemble (1) and $O^{\bar cc}_{1,3,5},O^{DD}_{1,3,4}$ for ensemble (2). $t_0=2$ is used for all data points.}
\end{table}

\begin{figure*}[htb]

\begin{center}
\includegraphics[height=4.cm,clip]{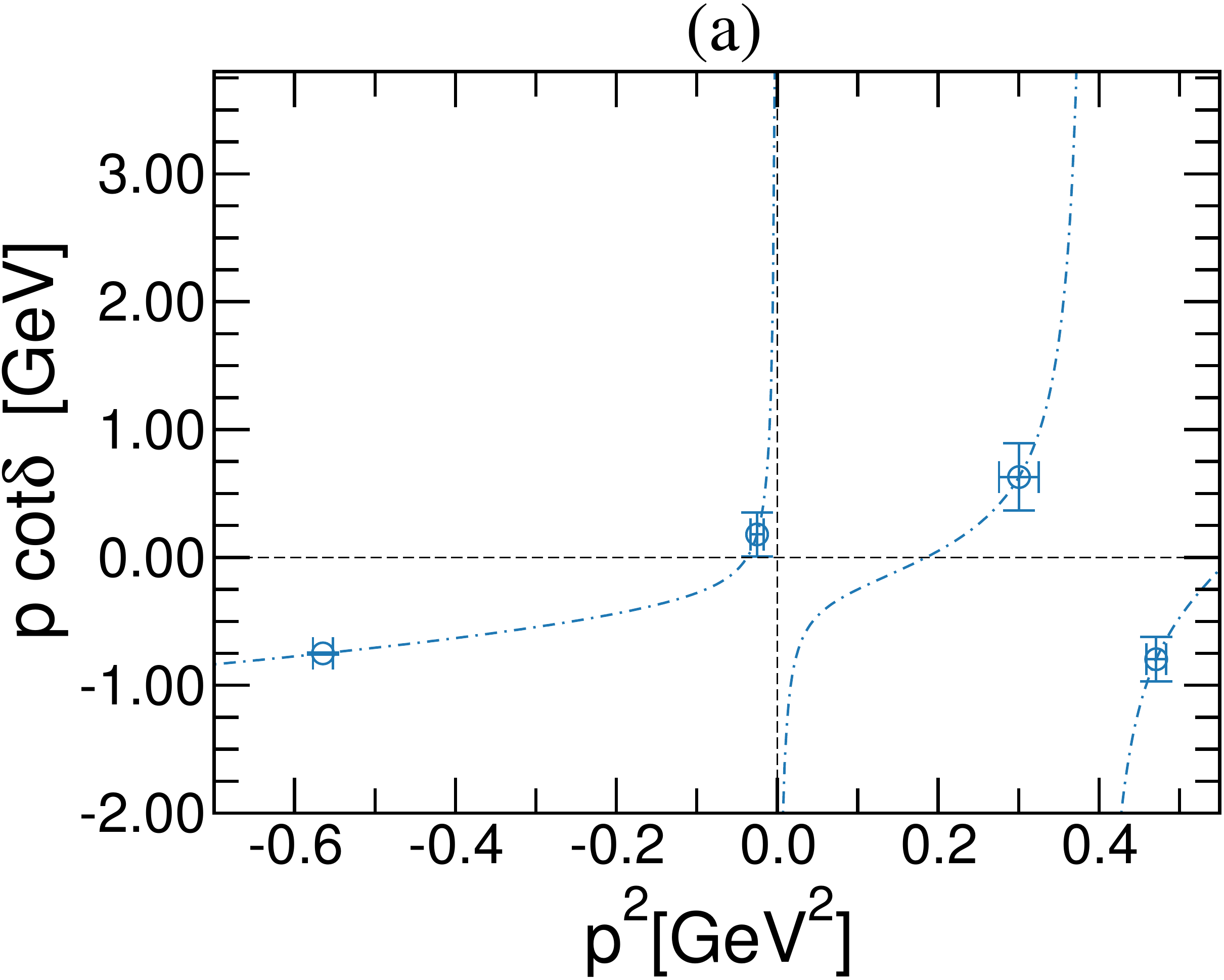}
\includegraphics[height=4.cm,clip]{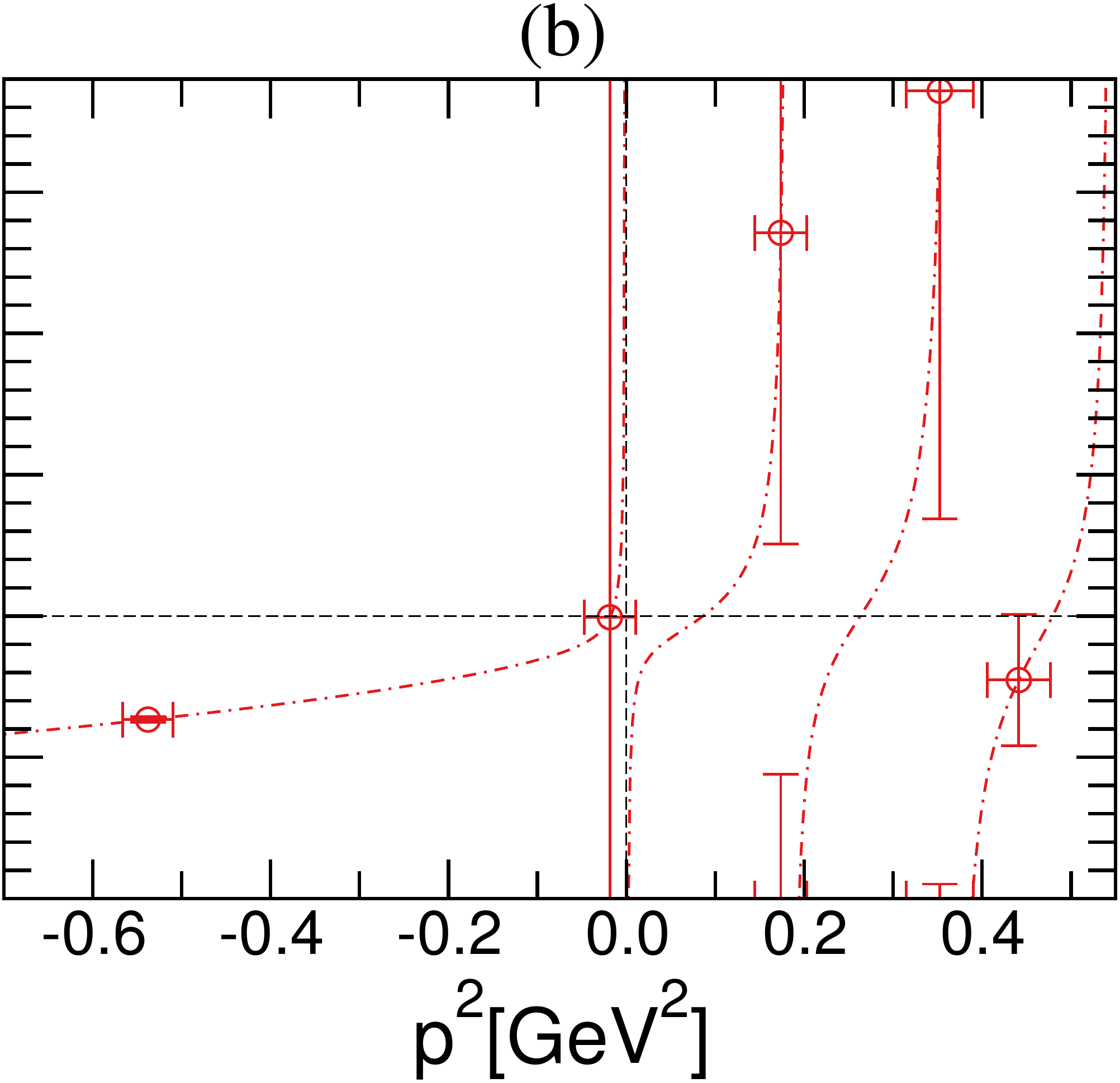}
\includegraphics[height=4.cm,clip]{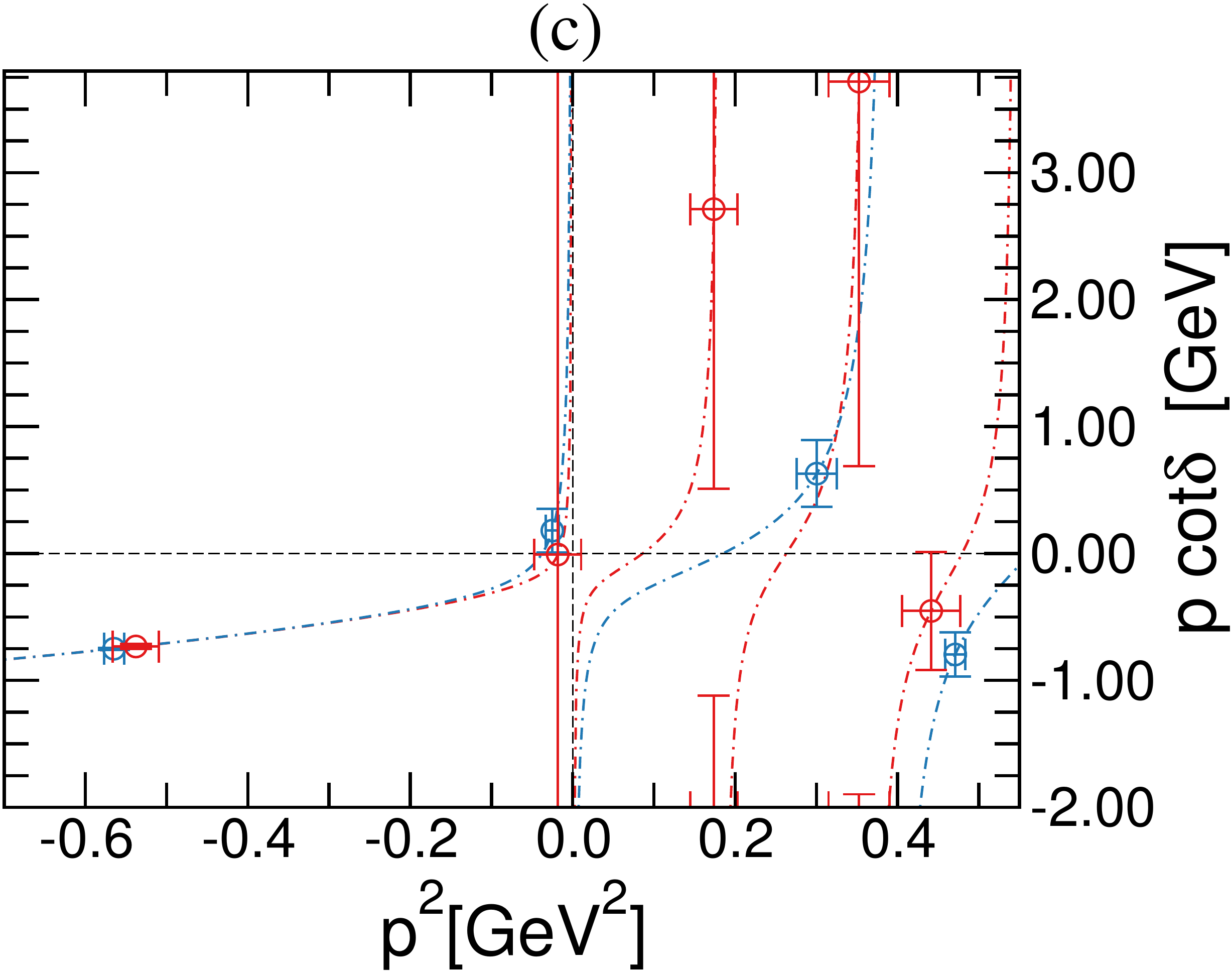}\\
\includegraphics[height=4.cm,clip]{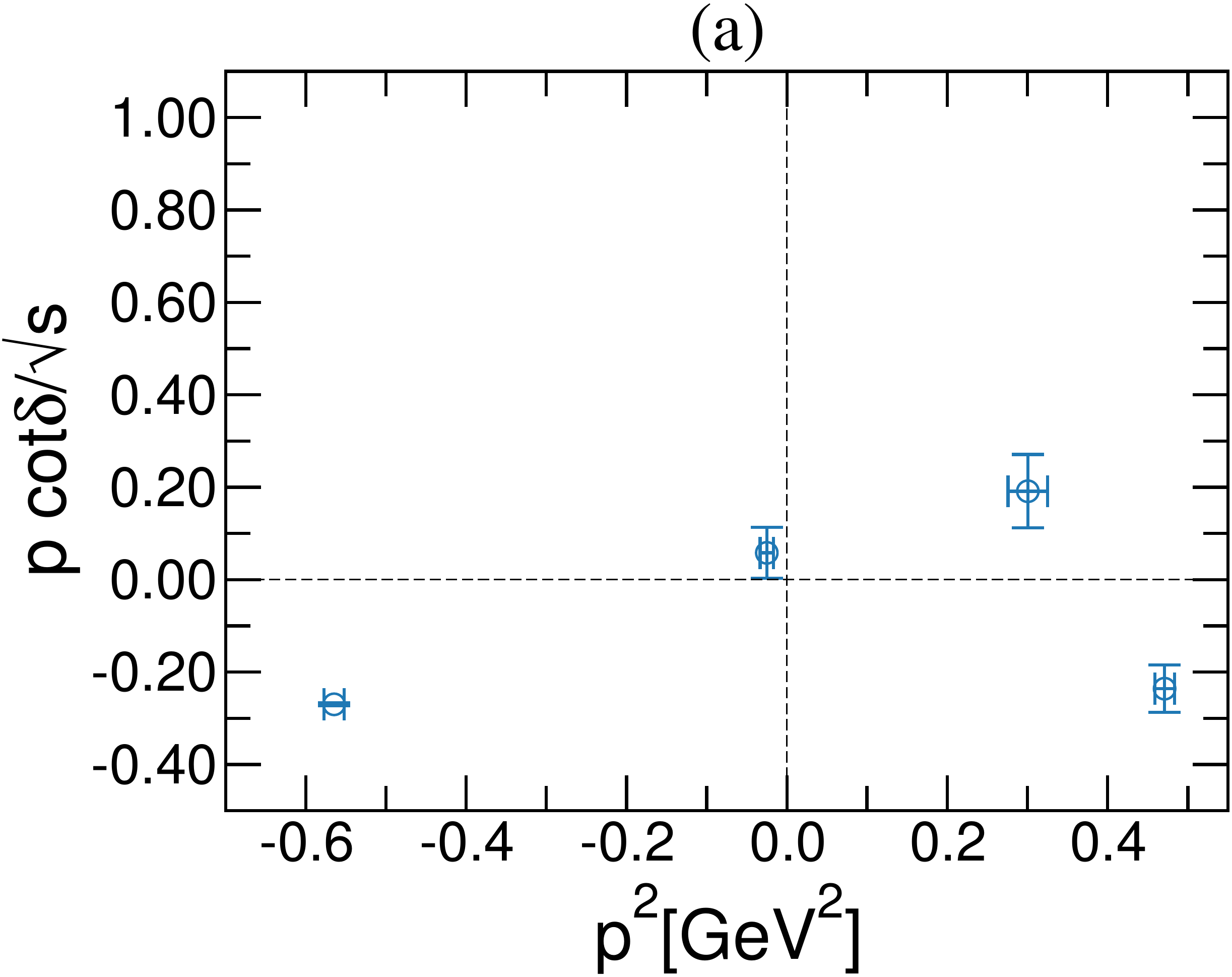}
\includegraphics[height=4.cm,clip]{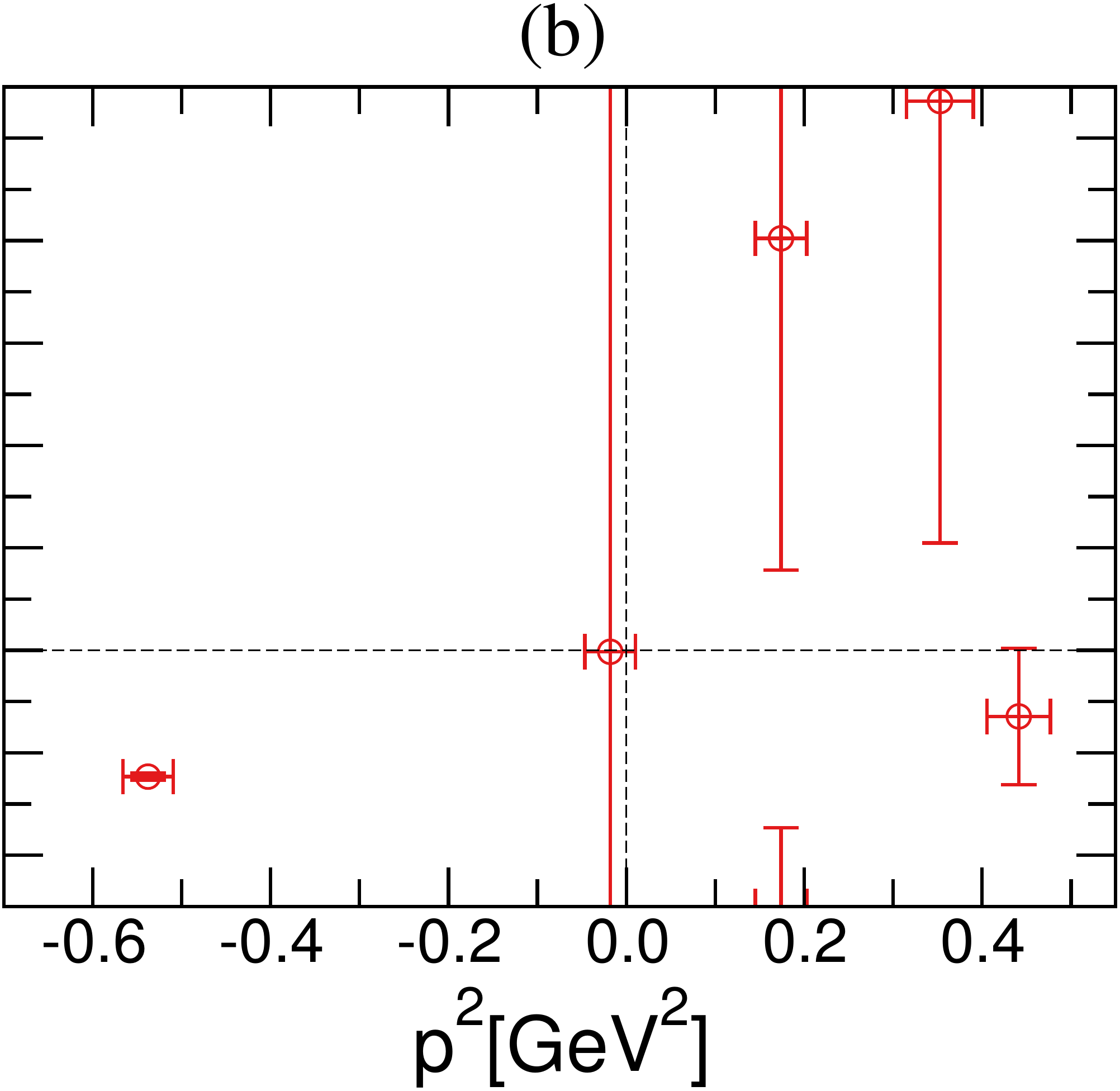}
\includegraphics[height=4.cm,clip]{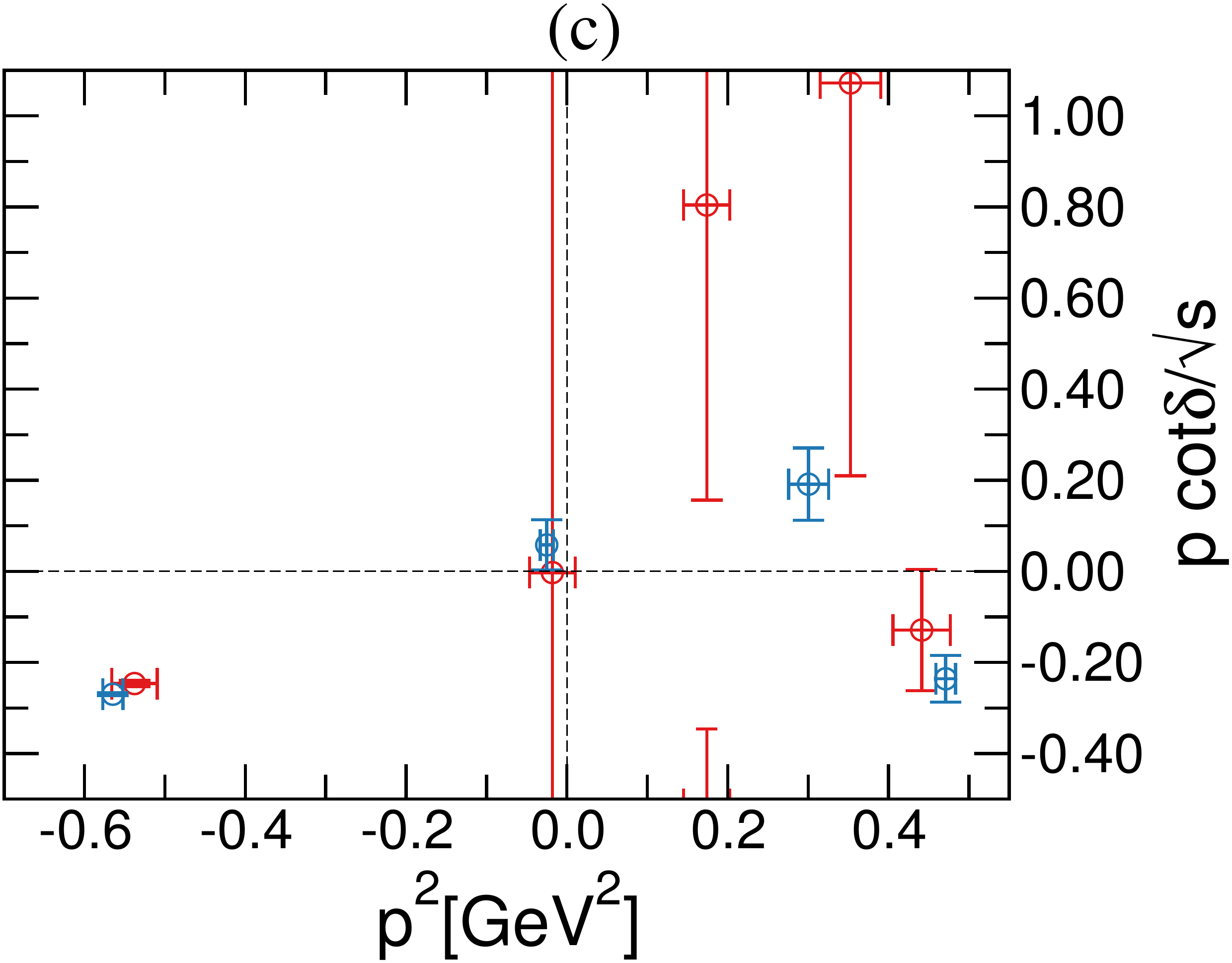} 
\end{center}
\caption{ We show  $p\cot\delta$ and $p\cot\delta/\sqrt{s}$ versus $p^2$ for $\bar D D$  scattering in $s$-wave, where $p$ denotes the momentum of the $D$ meson. The circles with (sizable) errors denote the lattice data, while the solid lines show   $p\cot\delta=2Z_{00}/(L\sqrt{\pi})$ according to L\"uscher's relation   (\ref{luscher}). When the momentum is compatible with the non-interacting  momentum $\mathbf{p}=2\pi \mathbf{q}/L$ ($\mathbf{q}\in N^3$), one has $\delta =0$ and $|\cot\delta|=\infty$, which is responsible for the huge errors on $p\cot\delta$ on ensemble (2).   }\label{fig:zeta_scalar} 
\end{figure*} 
    
\begin{figure*}[htb]
\begin{center}
\includegraphics[height=3.6cm,clip]{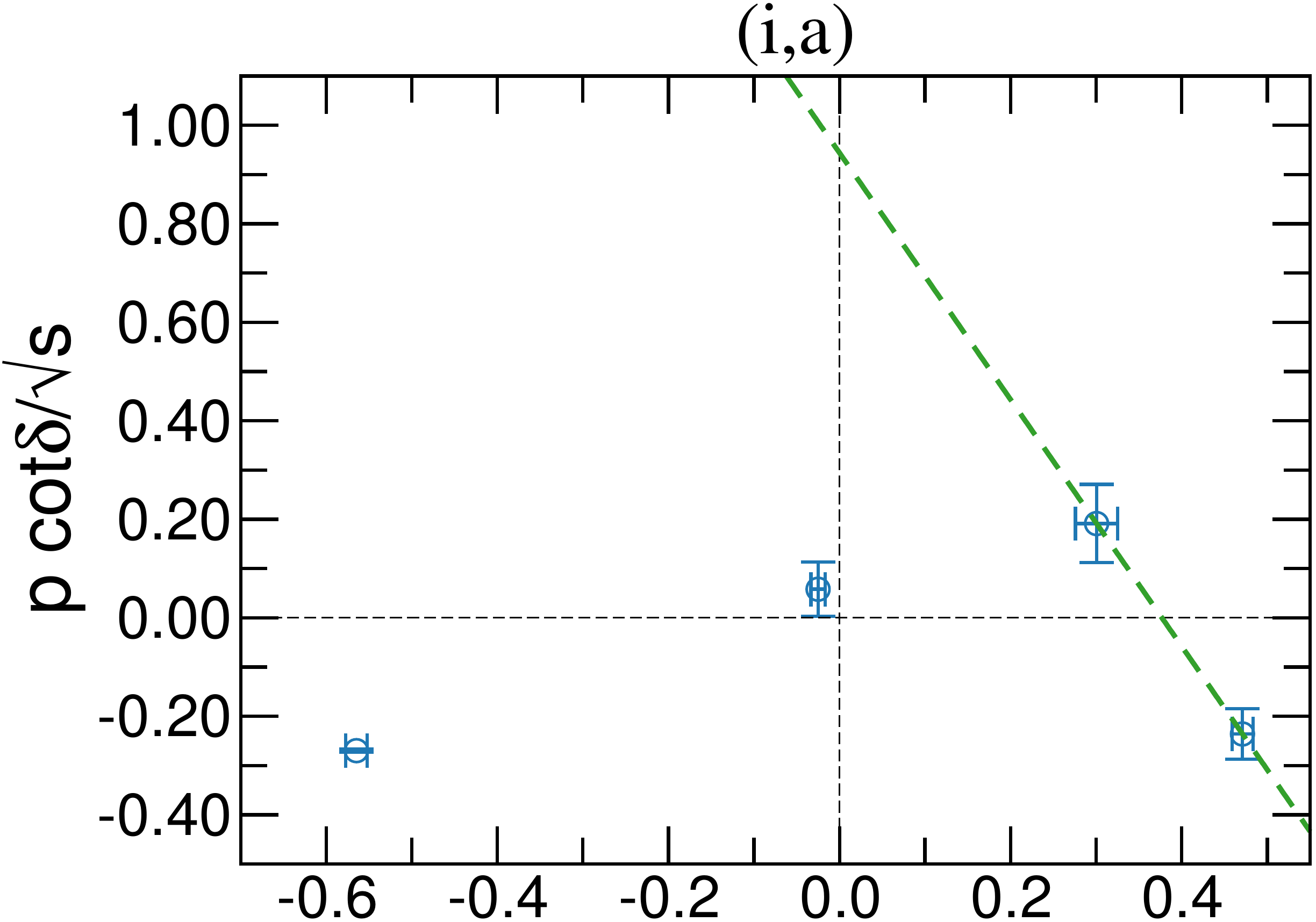}
\includegraphics[height=3.6cm,clip]{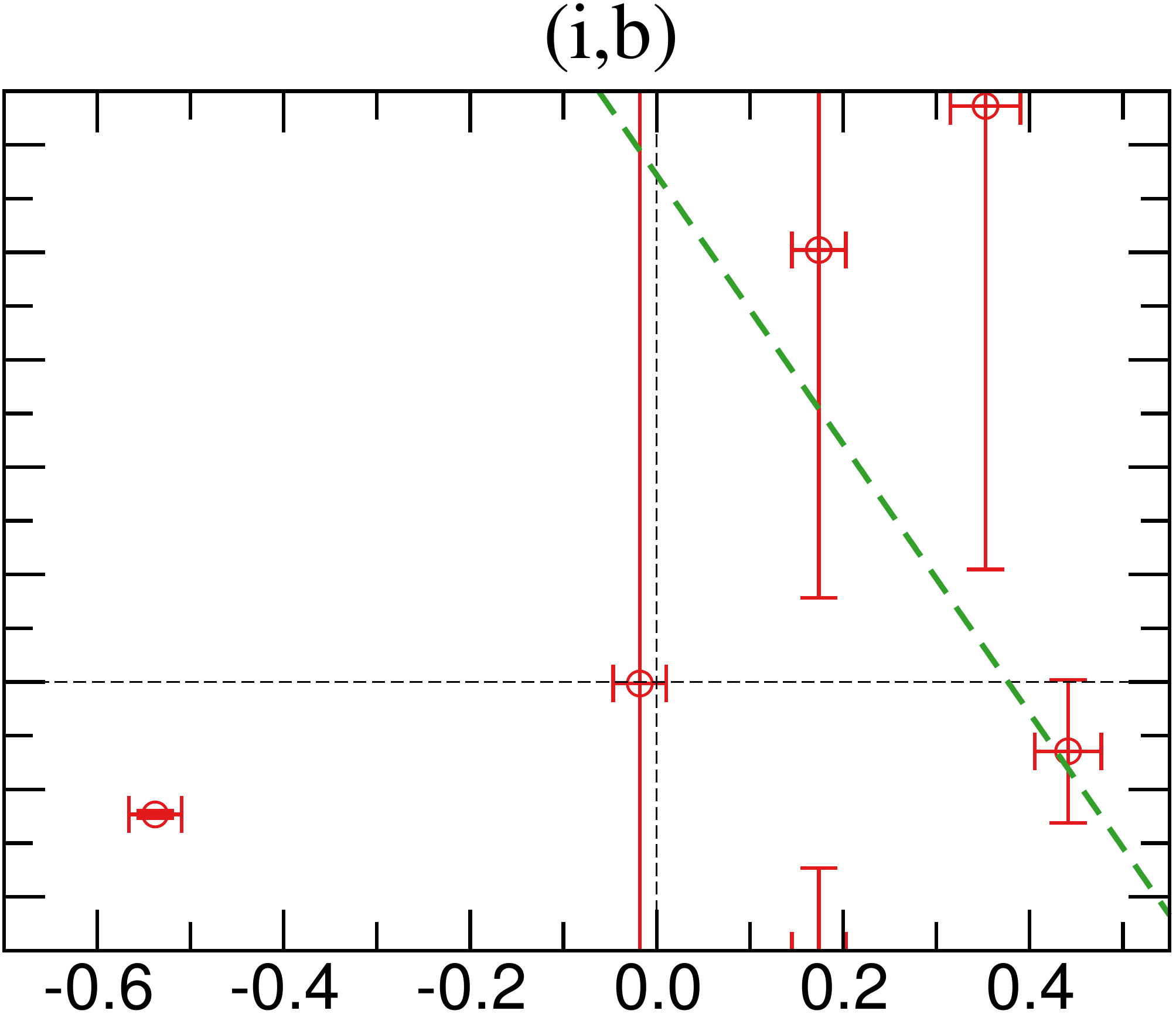}
\includegraphics[height=3.6cm,clip]{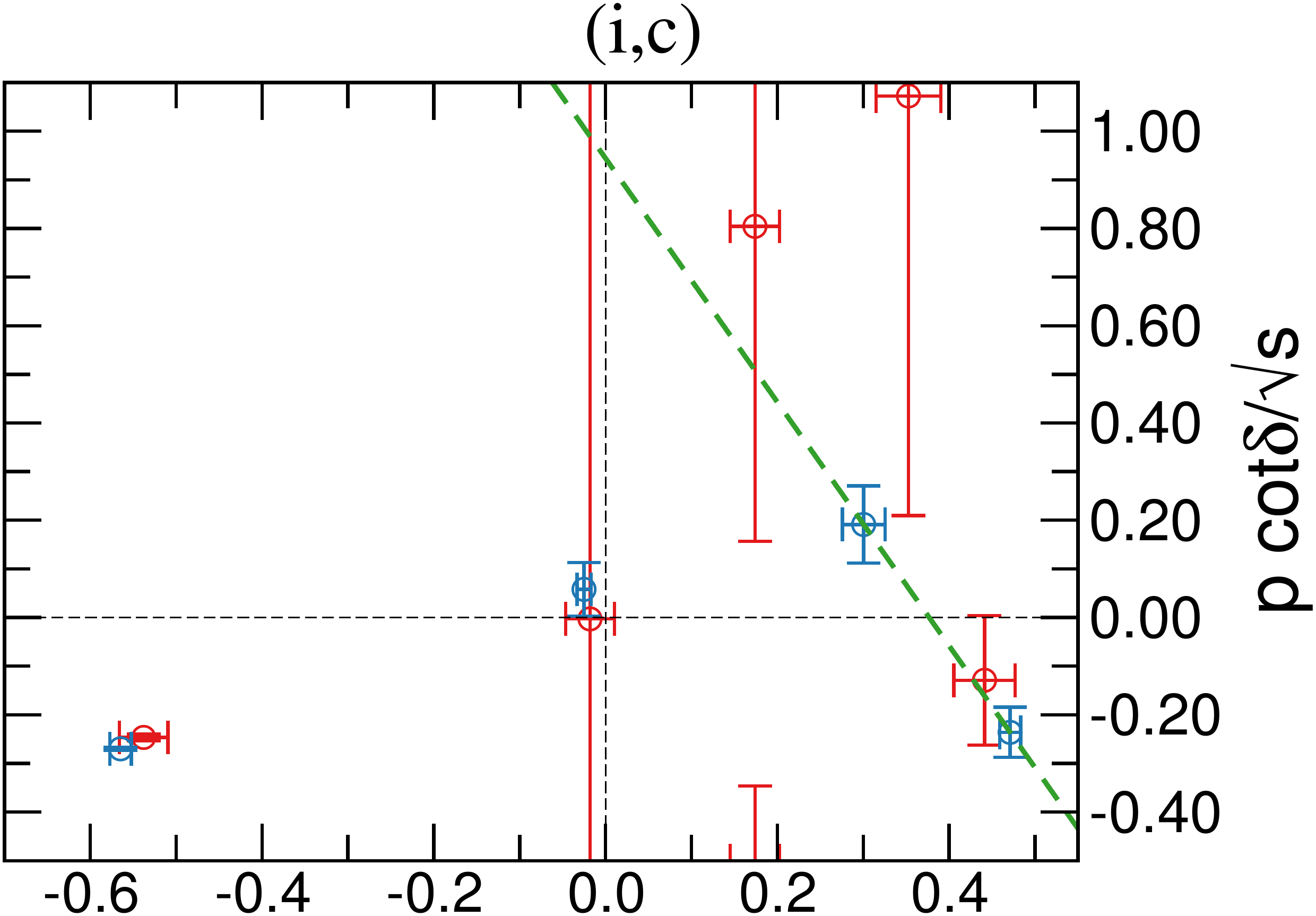}\\
\includegraphics[height=3.6cm,clip]{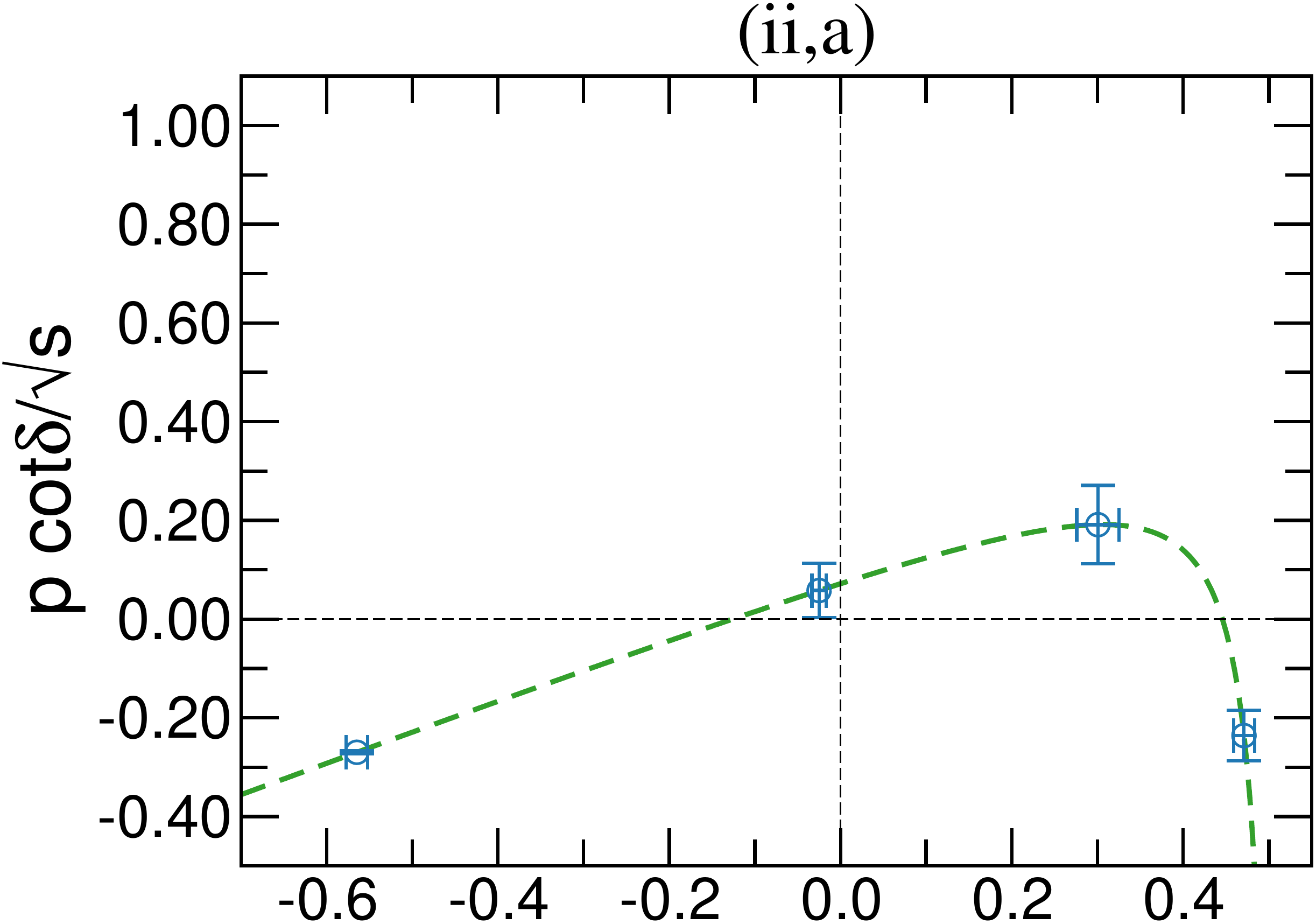}
\includegraphics[height=3.6cm,clip]{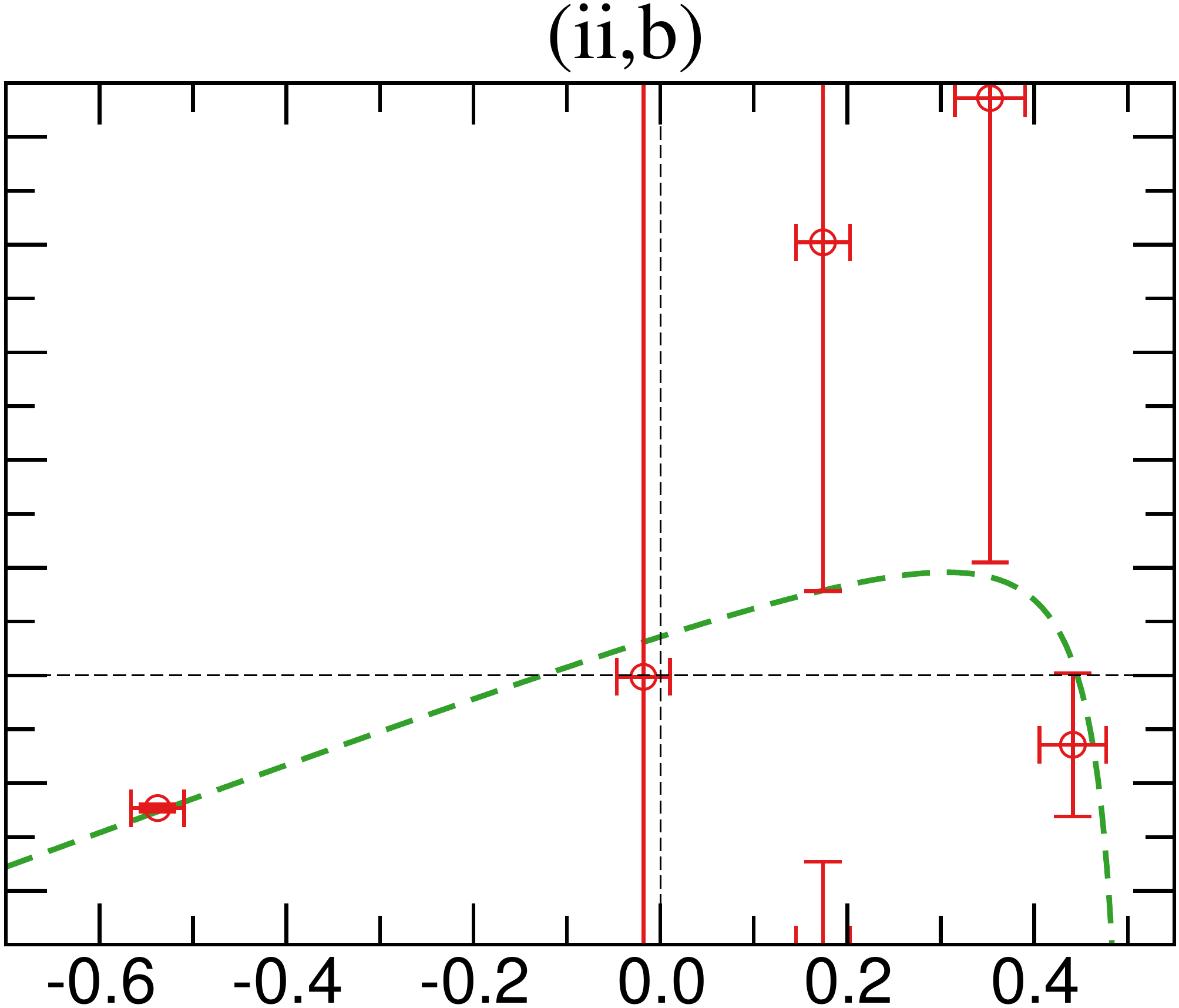}
\includegraphics[height=3.6cm,clip]{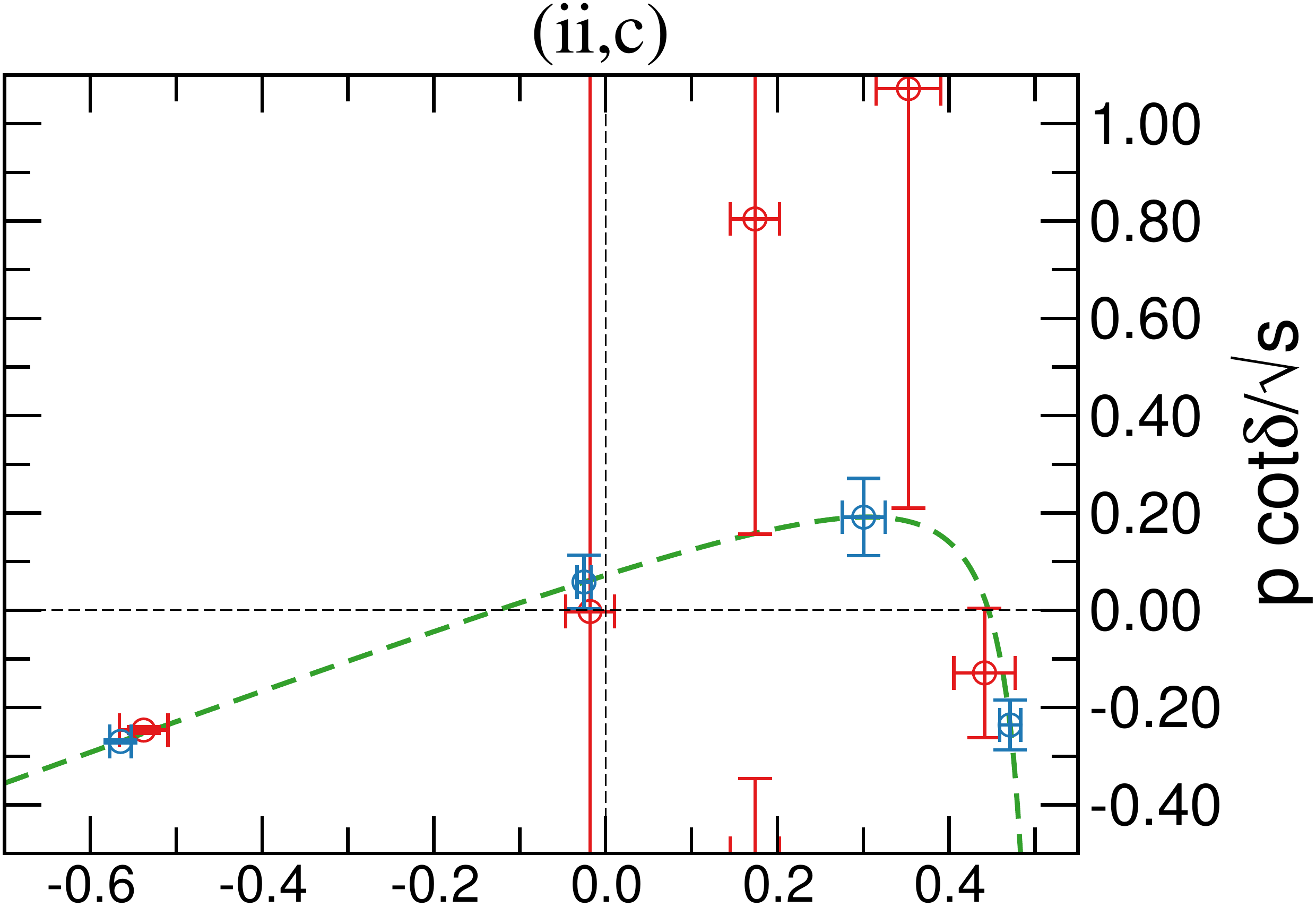}\\
\includegraphics[height=3.6cm,clip]{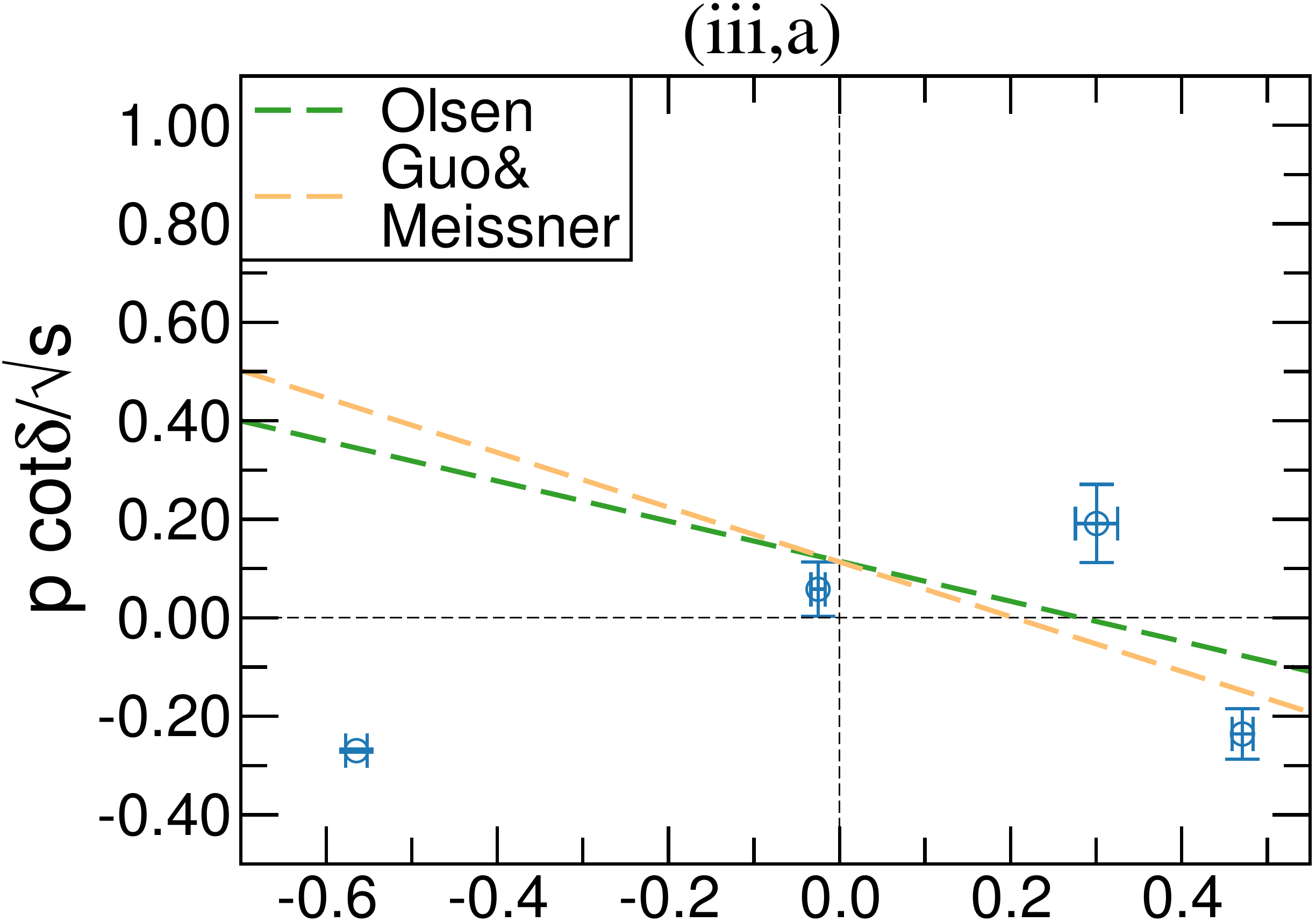}
\includegraphics[height=3.6cm,clip]{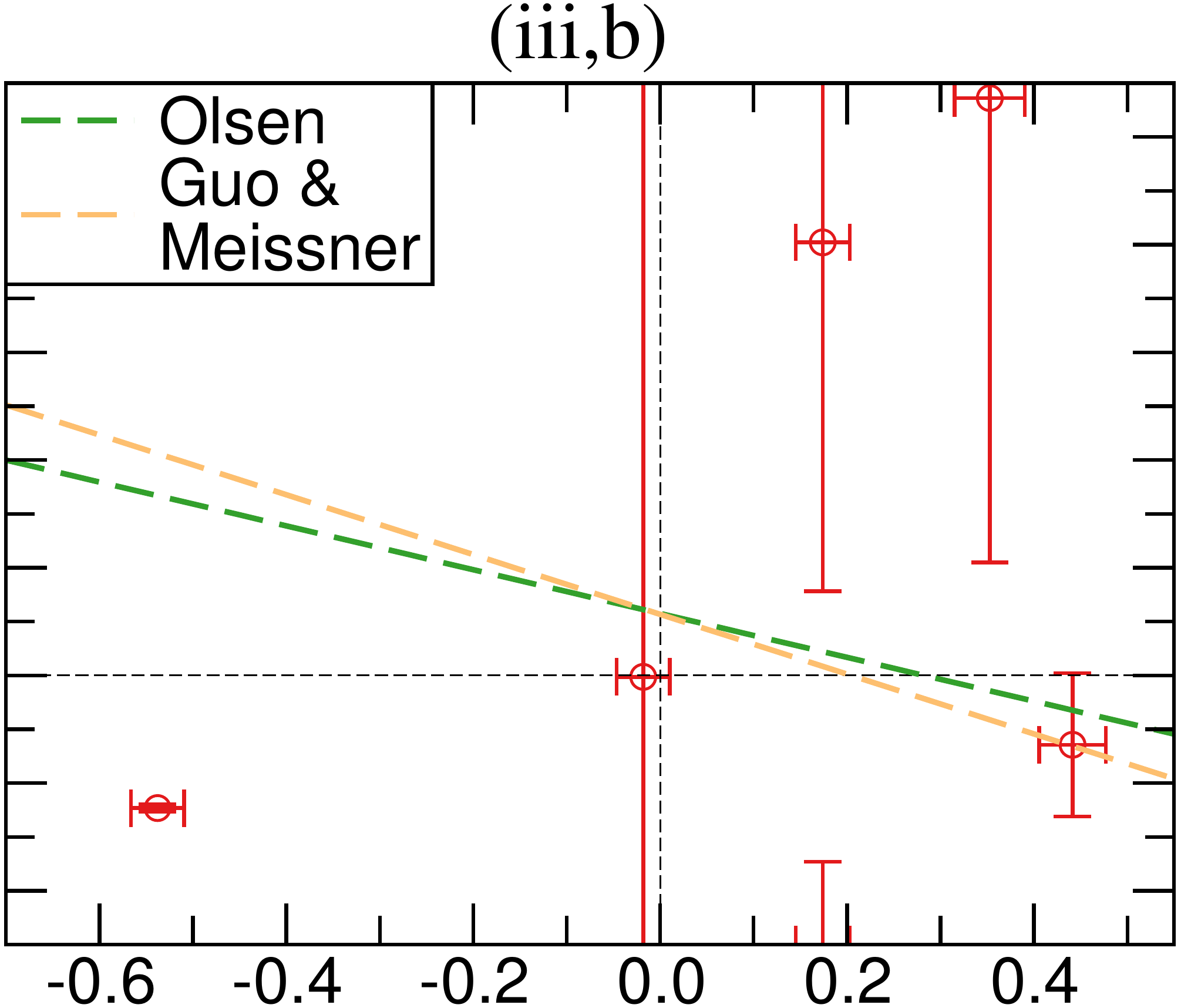}
\includegraphics[height=3.6cm,clip]{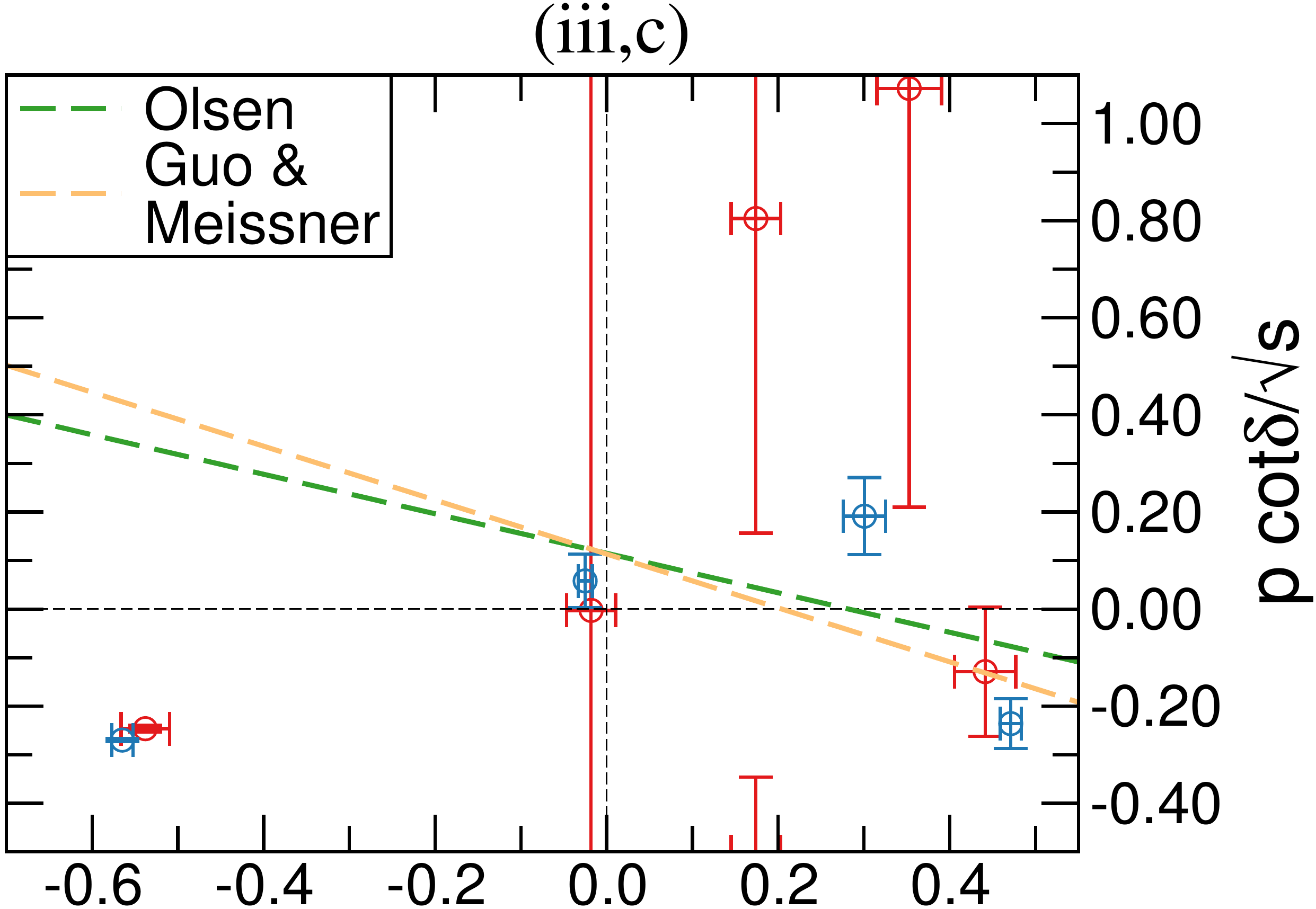}\\
\includegraphics[height=4.08cm,clip]{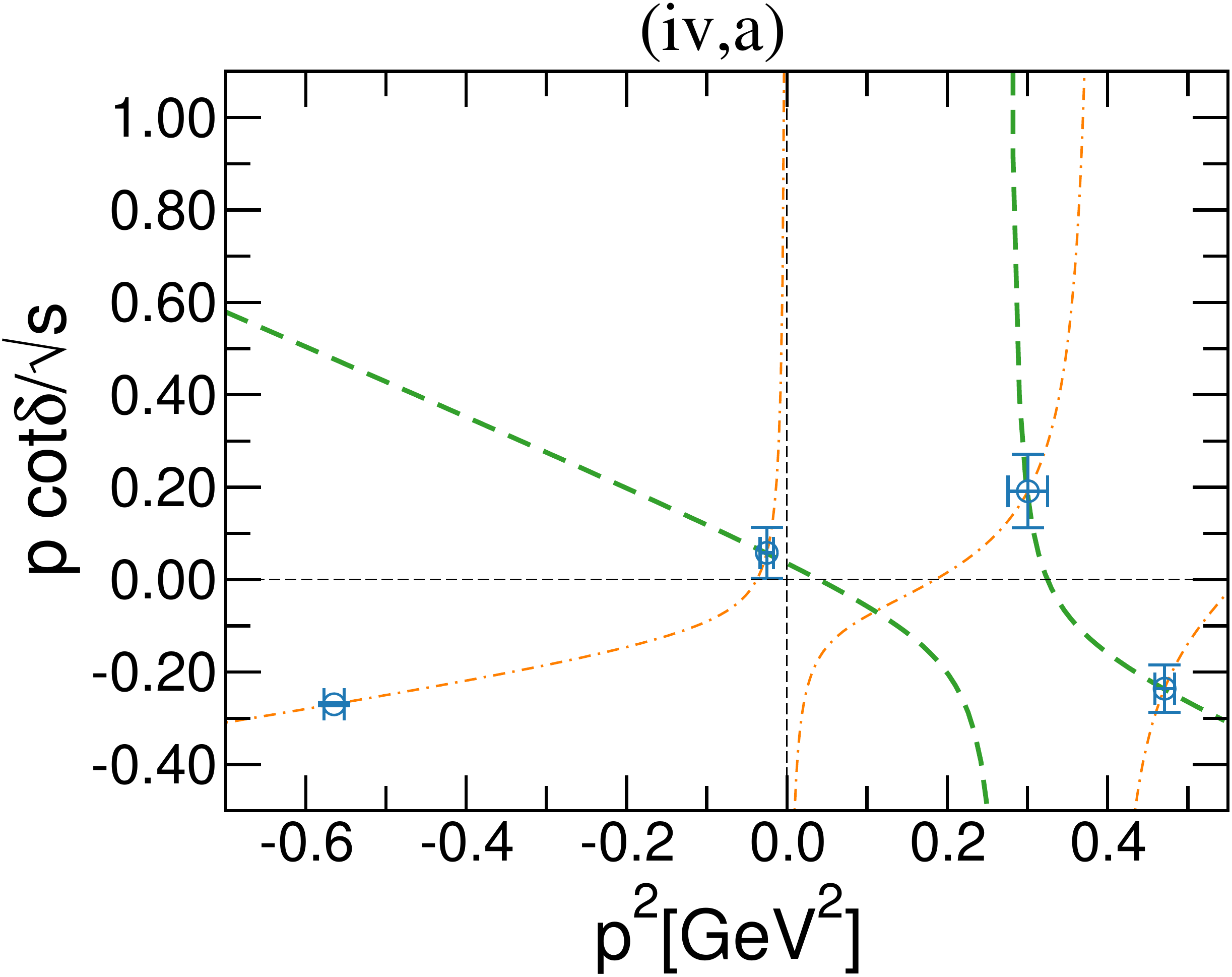}
\includegraphics[height=4.08cm,clip]{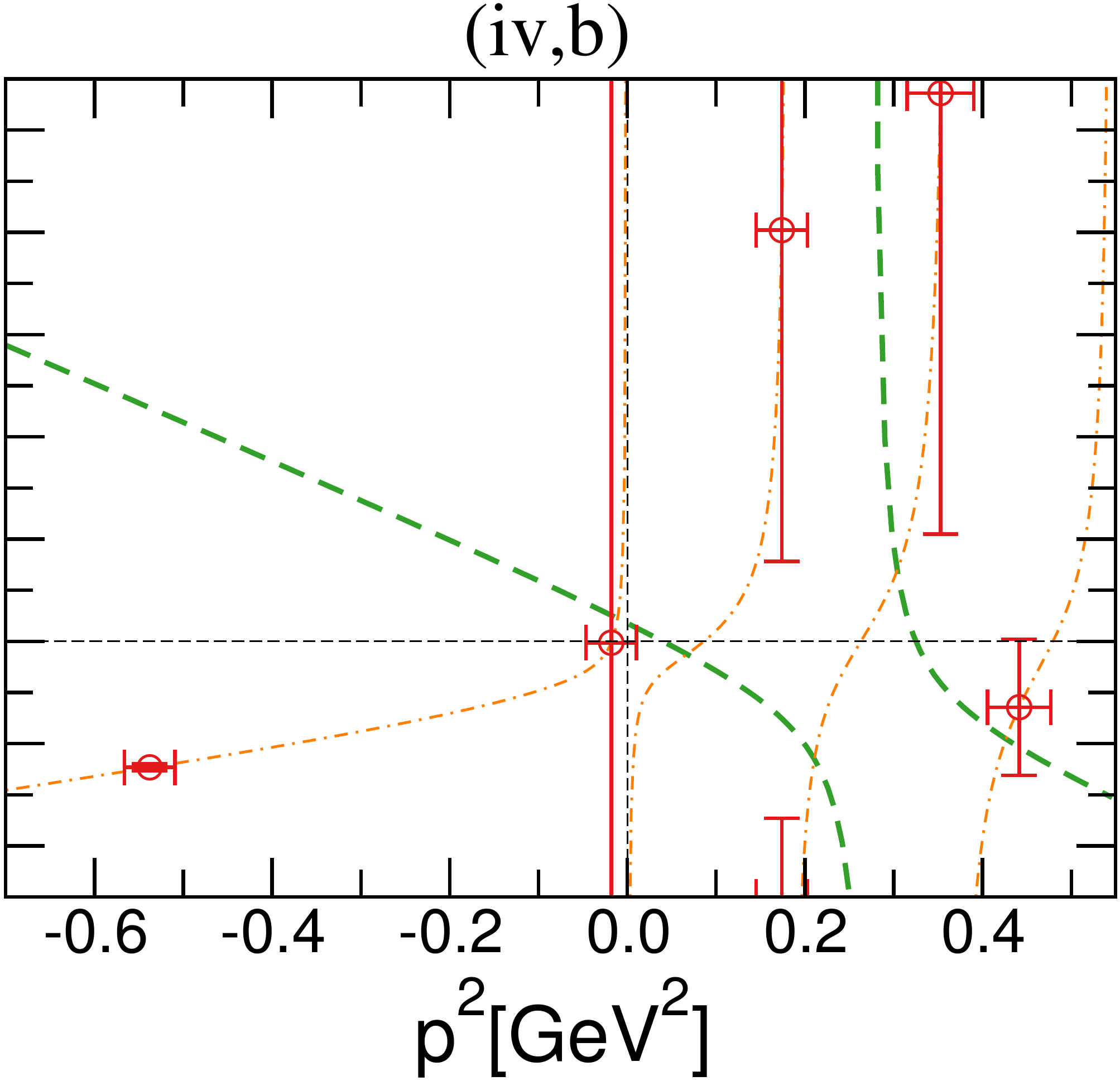}
\includegraphics[height=4.08cm,clip]{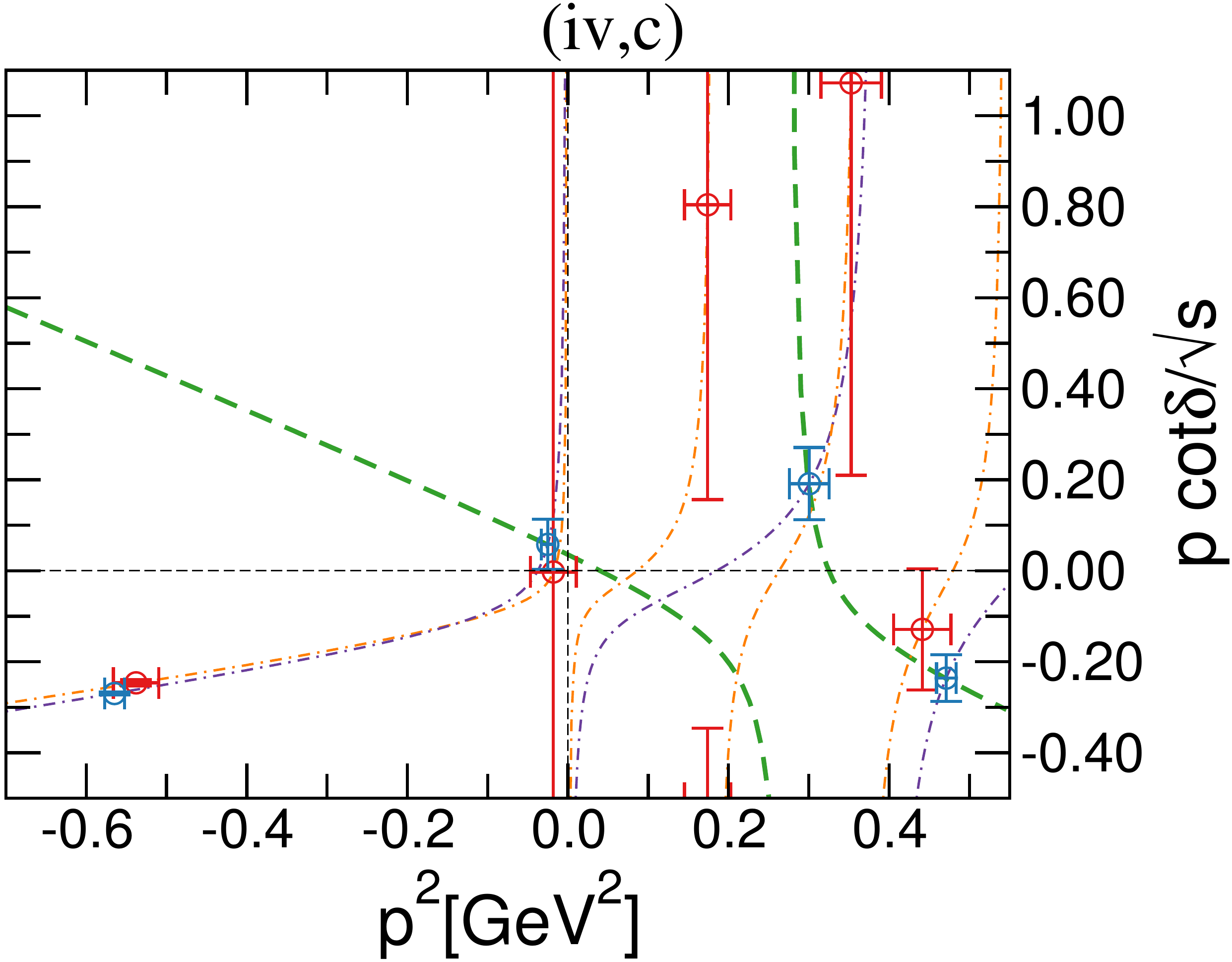}
 \end{center}
\caption{ The $p \cot\delta/\sqrt{s}$ versus $p^2$ for $\bar D D$  scattering in $s$-wave, where $p$ denotes the momentum of the $D$ meson.  The lattice data (blue and red circles) is confronted with $p \cot\delta/\sqrt{s}$ based on various hypothesis (dashed lines) described in Section \ref{sec:results_scalar} of the main text. The thin dot-dashed lines in the plots at the bottom denote $p\cot\delta=2Z_{00}/(L\sqrt{\pi})$ (\ref{luscher}). The left-hand column shows the results for ensemble (1), the middle column for ensemble (2) and the left-hand column is an overlay of both for comparison.}\label{fig:phase_scalar} 
\end{figure*}

We now study $\bar D D$   scattering  assuming that $J/\psi\,\omega$ channel is decoupled as argued in the previous paragraph. We did not include $\eta_c \eta$ interpolators in the correlation matrix assuming that they can be neglected. The energy shifts of the  extracted $E_n^{lat}$ with respect to $E^{n.i.}_{DD}$ give the size of the $s$-wave scattering phase shift $\delta$ according to (\ref{luscher}). On ensemble (1) we observe statistically significant energy shifts with respect to the dashed lines  in   Fig. \ref{fig:e_scalar}a. The energies yield $D$-meson momenta $p$ via $E^{lat}_n=2E_D(p)$, and the corresponding phase shifts $\delta(p)$ via Eq.  (\ref{luscher}).  These are provided for all levels in Table \ref{tab:e_scalar} and plotted in Figs. \ref{fig:zeta_scalar} and \ref{fig:phase_scalar}. 
    
The uncertainties on the energies $E_{n=2,3,4}$ are rather large for ensemble (2) and they are within errors compatible with non-interacting energies $E^{n.i.}_{DD}$    (\ref{ni}). This implies that we are not able to reliably determine the energy shifts, and the resulting errors on the scattering matrix  will be large, as illustrated in Fig. \ref{fig:zeta_scalar}. If  $E_n^{lat}\simeq E^{n.i.}_{DD}$ within errors, this implies $\delta\simeq 0$  modulo $\pi$ and $\cot\delta\simeq \pm \infty$ within errors.   The  extracted  $p\cot \delta$ from $n=2,3,4$ have large errors, which allow almost all   $p\cot \delta$ expect for small $|p\cot \delta|$. For $n=2,3,4$ we plot central values $f(p^2)$ with the ranges $[f(p^2-\sigma_{p^2}),f(p^2+\sigma_{p^2})]$ where $f=p\cot\delta$ or $p\cot\delta/\sqrt{s}$. The error for all other levels (on both ensembles and both channels) is the usual jack-knife.   
    
The resulting $p\cot \delta/\sqrt{s}$ for ensemble (1) has a puzzling  behavior and we are going to confront it with various hypothesis, collected in Fig. \ref{fig:phase_scalar}. The errors  of $p\cot \delta/\sqrt{s}$ on ensemble (2) are  large and do not allow reliable fits. We will still compare data from ensemble (2)  with fits that are based on ensemble (1), and plot them as function of $p^2$ on the same figure.  \vspace{6pt}

\noindent {\bf (i) A narrow resonance:  }
In the vicinity of a Breit-Wigner resonance  (\ref{amplitude}) one expects 
\begin{equation}
\label{fit_1_scalar}
\frac{p\cot\delta(s)}{\sqrt{s}}=\frac{4}{g^2}(p_{R}^2-p^2)~,\ \  \Gamma(s)=g^2\frac{p}{s}~
\end{equation}
and the zero gives the position of the resonance. 
The upper three points in Fig. \ref{fig:phase_scalar}a however do not fall onto one line, so our results cannot be reconciled with a single Breit-Wigner resonance in the region between $2m_D$ and $4~$GeV. The highest two points  support the existence of a narrow resonance between them and a linear fit (\ref{fit_1_scalar}) over two levels shown in Fig. \ref{fig:phase_scalar}(i,a)  renders  (\ref{mR}) 
\begin{eqnarray}
\label{fit_1_scalar_A}
m_R&=&3.966(20)~\mathrm{GeV}, \ g=1.26(18)~\mathrm{GeV},\\
\nonumber
p_R^{lat}&=&0.614(33)~\mathrm{GeV},\  \Gamma^{lat}=62(17)~\mathrm{MeV} ~.
\end{eqnarray} 
The scattering data from ensemble (1) therefore suggest the existence of a  yet unobserved  scalar state called $\chi_{c0}^\prime$. Note that its mass is within the range of the naive estimate (\ref{naive_scalar}). It has  a width $\Gamma^{lat}$ in our simulation, while the corresponding  width  $ g^2 \tilde p/m_R^2$ in experiment would be modified due to a different phase space via $\tilde p=[m_R^2/4-(m_{D}^{exp})^2]^{1/2}$, leading to
\begin{equation}
\Gamma^{exp}_{predict}=67(18)~\mathrm{MeV}~.
\end{equation}
We have assumed that $g$ and $m_R$ do not depend on the pion mass here.  
  It is unlikely that this state corresponds to $X(3915)$ since the $\bar D D$   decay channel was not observed for this state.  A narrow resonance is roughly consistent also with the result from ensemble (2) within huge errors (see Fig. \ref{fig:phase_scalar}(i,b)), however  there must be some additional interaction between $D$ and $\bar D$ near the threshold according to ensemble (1). \vspace{6pt}

\noindent {\bf (ii) A narrow resonance and a bound state $\mathbf{\chi_{c0}(1P)}$:}  
Our  next hypothesis assumes that  $\chi_{c0}(1P)$ represents a pole in $\bar D D$  scattering on the first Riemann sheet, leading to   $p\cot \delta \simeq i|p_B| i =-|p_B|$ at the position of the bound state.  The negative value of $p\cot\delta$ below threshold might be a possible reason why   $p\cot\delta$  at threshold is smaller than expected based on narrow resonance (\ref{fit_1_scalar},\ref{fit_1_scalar_A}). In this case the value of $p\cot \delta$ at threshold is influenced by the resonance and a bound state. To investigate this situation, we attempted several fits over all four levels on ensemble (1). A form that is motivated by our data 
\begin{equation}
\label{fit_2_scalar} 
\frac{p\cot\delta(s)}{\sqrt{s}}= A  + B~ p^2 + \frac{C}{p^2 - D}
\end{equation}
is presented in  Fig. \ref{fig:phase_scalar}(ii,a), where  $A=0.13(15)$, $B=0.66(18)$/GeV$^2$, $C=0.028(63)~$GeV$^2$ and $D=0.513(77)~$GeV$^2$ are obtained from the fit. This hypothesis supports a bound state at $p_B=i|p_B|$  which corresponds to a pole in $T$ (\ref{amplitude}) or equivalently  $\cot\delta(p_B)=i$, i.e. 
\begin{equation}
|p_B|= 0.7517(83)~\mathrm{GeV} \quad m_B=3.4224(27)~\mathrm{GeV}~.
\end{equation} 
The bound state is attributed to $\chi_{c0}(1P)$ and its mass is very close to the one obtained from the ground state energy.  The hypothesis also supports a narrow resonance at $p_R=0.668(35)~$GeV where function (\ref{fit_2_scalar})   crosses zero, and 
\begin{eqnarray}
\label{fit_2_scalar_B}
m_R&=&4.002(24)~\mathrm{GeV}, \ g=0.85(65)~\mathrm{GeV},\\
\nonumber
\Gamma^{lat}&=&30(45)~\mathrm{MeV}~, \Gamma^{exp}_{predict}=32(48)~\mathrm{MeV}~ .
\end{eqnarray}  
This is roughly consistent with the $\chi_{c0}^\prime$ in (\ref{fit_1_scalar_A}). This hypothesis based on ensemble (1) is consistent also with the result from ensemble (2) within huge errors in  Fig. \ref{fig:phase_scalar}(ii,b).  An interesting feature of this hypothesis is the large $p\cot\delta$ or equivalently small cross-section at $p^2\simeq D$, which corresponds to $\sqrt{s}\simeq 4.0~$GeV. This feature seems to be present also in the  experimental data from Belle  \cite{Olsen:2014maa} where a dip seems to appear  at similar invariant mass.  \\

\noindent  {\bf (iii) A broad resonance:}  
The broad resonances (\ref{guo_meissner},\ref{olsen}) proposed by Meissner\&Guo \cite{Guo:2012tv} or   Olsen \cite{Olsen:2014maa} are compared with our lattice data in Fig. \ref{fig:phase_scalar}(iii).  This shows  a Breit-Wigner shape  (\ref{fit_1_scalar}) with $p_R$ and $g$ extracted from the experimental data (\ref{guo_meissner},\ref{olsen}). Although they are roughly compatible with our scattering results near threshold, they cannot be reconciled with it in the region above threshold where our data indicates either a much narrower resonance or a more complicated situation not covered by our assumptions. \vspace{6pt}
 
\noindent  {\bf (iv) Two  resonances:}   
Since neither one narrow or one broad resonance  describe our scattering data near and above threshold, we next try an hypothesis with two elastic resonances
\begin{equation}
\label{fit_4_scalar}
\frac{p\cot\delta(s)}{\sqrt{s}}=\left[\frac{g_A^2}{4(p_{R_A}^2-p^2)}+\frac{g_B^2}{4(p_{R_B}^2-p^2)}\right]^{-1}~.
\end{equation} 
With this parametrisation there are two resonance poles in the scattering amplitude, separated by a zero. Figure \ref{fig:phase_scalar}(iv,a) shows an example with $g_{A}=2.1~$GeV, $p_{R_A}=0.23~$GeV,  $g_{B}=1.0~$GeV and  $p_{R_B}=0.57~$GeV  that is consistent with the  upper three scattering points for ensemble (1)\footnote{ These values are not obtained from a fit, but present  one example of four parameters, where  (\ref{fit_4_scalar}) is consistent with upper three scattering points.}. This hypothesis however  predicts   another energy level near $p^2\simeq 0.1~$GeV$^2$ where the  model (\ref{fit_4_scalar}) crosses with the L\"uscher curve. Another energy level is expected in the two-resonance scenario also according to naive reasoning that each resonance or bound state leads to a level in addition to $\bar D D$. Such an additional energy level at $p^2\simeq 0.1~$GeV$^2$  is not observed in ensemble (1) indicating that this hypothesis is not supported by our data. An analogous conclusion is reached when confronting this hypothesis with the data from ensemble (2):  the hypothesis predicts five energy levels  in the region $p^2=[-0.1,0.5]~$GeV and we observe four levels only.   
\section{Conclusions and outlook} 
We performed a lattice QCD simulation of $\bar{D}D$ scattering in $s$-wave and $p$-wave to study vector and scalar charmonium resonances on two rather different ensembles. This is an exploratory simulation and the first step towards determining the strong decay width of charmonium resonances above open charm threshold. Ensemble (1) has $N_f\!=\!2$ and $m_{\pi}=266$ MeV, while ensemble (2) has $N_{f}\!=\!2+1$  and $m_{\pi}=156$ MeV. Several $\bar cc$ and $D\bar D$ interpolating fields were used in both channels, where the  (stochastic) distillation method  was used to evaluate the Wick contractions. Our analysis relies on the assumption that looking at elastic scattering  in a single channel ($\bar{D}D$) is a good approximation.

In the vector channel, the well known $\psi(3770)$ resonance is present just above $\bar{D}D$ threshold with $Br^{exp}[\psi(3770)\to D\bar D]=93\pm 9\%$. We assume that the $D\bar D$ scattering is elastic in this energy region and determine  the phase shift for $D\bar D$ scattering in $p$-wave using the L\"uscher formalism.  The Breit-Wigner fit  is performed in vicinity of the $\psi(3770)$ to obtain its resonance mass at $3.784(7)(10)~$GeV and $3.786(56)(10)~$GeV for ensembles (1) and (2), respectively.  Our determination of its decay width might be affected by the $\Psi(4040)$ on ensemble (1). Ensemble (2) does not suffer from this issue, and the determination of the resonance parameters is more reliable, but its statistical accuracy is poor. The resulting spectrum in the vector channel, including also $J/\psi$ and $\psi(2S)$, is compared to experiment in Figure \ref{fig:summary_vector}.  
This work presents a step towards a determination of the $\psi(3770)$ resonance parameters from lattice QCD. Improvement of  the results for this resonance in future lattice studies will need consideration of multiple volumes and momentum frames, further scattering channels and higher statistics.

In the scalar channel, only the ground state $\chi_{c0}(1P)$ is understood and there is no commonly accepted candidate  for its first excitation  $\chi_{c0}(2P)$.  Guo \& Meissner \cite{Guo:2012tv} as well as Olsen \cite{Olsen:2014maa} argued that the higher lying $X(3915)$ can probably not be identified with the $\chi_{c0}(2P)$. They suggest that a broad structure observed in the $D\bar D$  invariant mass  represents $\chi_{c0}(2P)$. This posed a particular motivation to extract the phase shift for $D\bar D$ scattering in $s$-wave in the present work.  The resulting scattering data on the ensemble with $m_\pi=156~$MeV is unfortunately  noisy. The simulation at $m_\pi=266~$MeV renders the scattering phase shift only at a few values of the $D\bar D$ invariant mass, which also does not allow a clear answer to the puzzles in this channel.  We obtain the $\chi_{c0}(1P)$ and our data provides an indication for a yet-unobserved narrow resonance slightly below $4~$GeV with $\Gamma[\chi_{c0}^\prime \to D\bar D]$ below $100~$MeV.  A scenario with this narrow resonance and a pole in the $D\bar D$ scattering matrix at $\chi_{c0}(1P)$ agrees  with the energy-dependence of our phase shift. We discussed three other scenarios: just one narrow resonance, just one broad resonance  (proposed in  Guo \& Meissner \cite{Guo:2012tv} and Olsen \cite{Olsen:2014maa}), or one narrow and one broad resonance. None of these scenarios agree with our current data in the whole energy region probed, however we can not currently exclude these possibilities. For the scalar channel this leaves us with a situation where puzzles remain, both from theory and experiment.  To clarify the situation, further experimental and lattice QCD efforts are required to map out the $s$-wave $D\bar D$ scattering in more detail. Including further coupled channels in the future would be useful to relax the model assumptions made in our current study.

\acknowledgments
We  thank Anna Hasenfratz and the PACS-CS for providing the gauge configurations and Martin L\" uscher for making his DD-HMC software available. The calculations were performed on computing clusters at TRIUMF, the University of Graz (NAWI Graz) and at Jozef Stefan Institute. This work is supported in part by the Austrian Science Fund FWF: I1313-N27, by the Slovenian Resarch Agency ARRS project N1-0020. Fermilab is operated by Fermi Research Alliance, LLC under Contract No. De-AC02-07CG11359 with the United States Department of Energy. S.P. acknowledges support from U.S. Department of Energy contract DE-AC05-06OR23177, under which Jefferson Science Associates, LLC, manages and operates Jefferson Laboratory.  


\providecommand{\href}[2]{#2}\begingroup\raggedright
\endgroup

\end{document}